\documentclass[letter]{aa} 
\usepackage{graphicx}
\usepackage{color}
\usepackage{txfonts}
\usepackage{hyperref}
\usepackage{lscape}
\usepackage{natbib}
\usepackage{amsmath}
\usepackage{amsfonts}
\usepackage{wasysym}
\usepackage{graphics}
\usepackage{times}
\usepackage{parskip}
\usepackage{pdflscape}
\usepackage{geometry}
\usepackage{marginnote}
\usepackage{multicol}
\usepackage{soul}
\usepackage{lmodern}
\usepackage{rotate}

\usepackage{natbibspacing}
    \renewcommand{\dbltopfraction}{0.9} 
\newfont{\tlx}{cmssdc10 scaled 600}
\newfont{\mlx}{cmssdc10 scaled 770}
\newfont{\olx}{cmssdc10 scaled 840}
\newfont{\nlx}{cmssdc10 scaled 900}
\newfont{\mfont}{cmssdc10 scaled 1100}
\newfont{\rfont}{cmti12 scaled 840}
\newfont{\hvss}{cmssdc10 scaled 1540}
\definecolor{myblue1}{rgb}{0.0,0.604,0.831} 
\definecolor{myblue2}{rgb}{0.0,0.49,0.6745}
\definecolor{myblue3}{rgb}{0.0156,0.4078,0.9921}
\definecolor{myblue4}{rgb}{0.0,0.44,0.87}
\definecolor{myred1}{rgb}{0.529,0.019,0.017}
\definecolor{mycyan}{rgb}{0.63921569,0.0,0.48235294}
\definecolor{mygreen}{rgb}{0.3568,0.54902,0.2549}
\definecolor{applegreen}{rgb}{0.55, 0.71, 0.0}
\definecolor{cadmiumgreen}{rgb}{0.0, 0.42, 0.24}
\definecolor{lila}{rgb}{0.8,0.333,1.0}
\definecolor{reffig}{rgb}{0.0,0.6784,0.93725}

\newcommand{\brem}[1]{\textcolor{black}{\nlx #1}}

\newcommand{\PutLabel}[3]{\put(#1,#2){#3}}

\hypersetup{
    pdftoolbar=true,				
    pdfmenubar=true,				
    pdffitwindow=false,			
    pdfstartview={FitH},			
    pdftitle={title},				
    pdfauthor={Author},			
    pdfsubject={Subject},			
    pdfcreator={Author},			
    pdfproducer={Author},			
    pdfkeywords={Population \& Evolutionary} ,        
    pdfnewwindow=true,			
    colorlinks=true,				
    linkcolor=cyan,				
    citecolor=myblue4,				
    filecolor=cyan,				
    urlcolor=cyan				
}
\def\FD{\sc Fado\rm}
\def\P3D{\sc Porto3D\rm}
\def\starlight{\sc Starlight\rm}
\def\SL{\sc Starlight\rm}

\def\?{{\bf\color{red}?}}

\def\reff{$R_{\rm eff}$}

\def\ha{H$\alpha$}
\def\hb{H$\beta$}

\def\msun{$\mathrm{M}_{\odot}$}
\def\zsun{$\mathrm{Z}_{\odot}$}

\def\D4000{$D_{4000}$}

\newcommand{\sbb}{mag/$\sq\arcsec$}
\newcommand{\lbb}{1/s\,$\sq\arcsec$}

\def\mstar{${\cal M}_{\star}$}

\def\msb{$\mu_{\star}$}

\def\ewha{EW(H$\alpha$)}

\def\lyc{Ly$_{\mathrm{c}}$}

\def\mustar{$\mu_{\star}$}

\def\tln2ha{$\log$([N\,{\sc ii}]${\scriptstyle 6584}$/H$\alpha$)}
\def\tlo3hb{$\log$([O\,{\sc iii}]${\scriptstyle 5007}$/H$\beta$)}
\newcommand{\tref}[1]{\textcolor{myblue4}{#1}}

\def\pegase{{\sc P\'egase}}
\def\zet{$z$}
\newcommand\btab[5]{\begin{table*}[#1]\label{#3}{\parbox{#4}{\caption{#2}}\rule[-0.5ex]{0cm}{0.5ex} }
\begin{tabular*}{#4}{#5} \label{#3} }

\def\lya{Ly$\alpha$}

\def\kc{{\it k}}

\def\cmod{{\sc Cmod}}

\def\la{\lesssim}
\def\ga{\gtrsim}
\def\lyc{Ly{\sc c}}
\def\NLyC{${\cal N}_{\rm LyC}$}
\def\exstars{$A^{\star}_{V}$}
\def\exneb{$A^{\rm neb}_{V}$}
\def\ne{NE}

\newcommand{\imCMOD}[8]{
\begin{center}
\begin{figure*}
\begin{picture}(200,130)
\put(0,67){\includegraphics[height=6.2cm]{./simimages/PNG_#1/#1_#4}}
\put(62,70){\includegraphics[height=5.6cm]{./simimages/PNG_#1/#1_#5}}
\put(124,70){\includegraphics[height=5.6cm]{./simimages/PNG_#1/#1_#6}}
\put(0,49.5){\includegraphics[height=1.6cm]{./simimages/PNG_#1/#1_#8_mz0.0.png}}
\put(15.2,49.5){\includegraphics[height=1.6cm]{./simimages/PNG_#1/#1_#8_mz0.15.png}}
\put(30.4,49.5){\includegraphics[height=1.6cm]{./simimages/PNG_#1/#1_#8_mz0.35.png}}
\put(45.6,49.5){\includegraphics[height=1.6cm]{./simimages/PNG_#1/#1_#8_mz0.7.png}}
\put(60.8,49.5){\includegraphics[height=1.6cm]{./simimages/PNG_#1/#1_#8_mz0.9.png}}
\put(76,49.5){\includegraphics[height=1.6cm]{./simimages/PNG_#1/#1_#8_mz1.3.png}}
\put(91.2,49.5){\includegraphics[height=1.6cm]{./simimages/PNG_#1/#1_#8_mz1.6.png}}
\put(106.4,49.5){\includegraphics[height=1.6cm]{./simimages/PNG_#1/#1_#8_mz2.0.png}}
\put(121.6,49.5){\includegraphics[height=1.6cm]{./simimages/PNG_#1/#1_#8_mz2.5.png}}
\put(136.8,49.5){\includegraphics[height=1.6cm]{./simimages/PNG_#1/#1_#8_mz3.0.png}}
\put(152,49.5){\includegraphics[height=1.6cm]{./simimages/PNG_#1/#1_#8_mz4.0.png}}
\put(167.2,49.5){\includegraphics[height=1.6cm]{./simimages/PNG_#1/#1_#8_mz5.0.png}}
\PutLabel{1}{62}{\nlx \textcolor{#3}{0.00}}
\PutLabel{16.2}{62}{\nlx \textcolor{#3}{0.15}}
\PutLabel{31.4}{62}{\nlx \textcolor{#3}{0.35}}
\PutLabel{46.6}{62}{\nlx \textcolor{#3}{0.70}}
\PutLabel{61.8}{62}{\nlx \textcolor{#3}{0.90}}
\PutLabel{77}{62}{\nlx \textcolor{#3}{1.30}}
\PutLabel{92.2}{62}{\nlx \textcolor{#3}{1.60}}
\PutLabel{107.4}{62}{\nlx \textcolor{#3}{2.00}}
\PutLabel{122.6}{62}{\nlx \textcolor{#3}{2.50}}
\PutLabel{137.8}{62}{\nlx \textcolor{#3}{3.00}}
\PutLabel{153}{62}{\nlx \textcolor{#3}{4.00}}
\PutLabel{168.2}{62}{\nlx \textcolor{#3}{5.00}}
\put(0,33){\includegraphics[height=1.6cm]{./simimages/PNG_#1/#1_F090WF150W_mz0.0.png}}
\put(15.2,33){\includegraphics[height=1.6cm]{./simimages/PNG_#1/#1_F090WF150W_mz0.15.png}}
\put(30.4,33){\includegraphics[height=1.6cm]{./simimages/PNG_#1/#1_F090WF150W_mz0.35.png}}
\put(45.6,33){\includegraphics[height=1.6cm]{./simimages/PNG_#1/#1_F090WF150W_mz0.7.png}}
\put(60.8,33){\includegraphics[height=1.6cm]{./simimages/PNG_#1/#1_F090WF150W_mz0.9.png}}
\put(76,33){\includegraphics[height=1.6cm]{./simimages/PNG_#1/#1_F090WF150W_mz1.3.png}}
\put(91.2,33){\includegraphics[height=1.6cm]{./simimages/PNG_#1/#1_F090WF150W_mz1.6.png}}
\put(106.4,33){\includegraphics[height=1.6cm]{./simimages/PNG_#1/#1_F090WF150W_mz2.0.png}}
\put(121.6,33){\includegraphics[height=1.6cm]{./simimages/PNG_#1/#1_F090WF150W_mz2.5.png}}
\put(136.8,33){\includegraphics[height=1.6cm]{./simimages/PNG_#1/#1_F090WF150W_mz3.0.png}}
\put(152,33){\includegraphics[height=1.6cm]{./simimages/PNG_#1/#1_F090WF150W_mz4.0.png}}
\put(167.2,33){\includegraphics[height=1.6cm]{./simimages/PNG_#1/#1_F090WF150W_mz5.0.png}}
\PutLabel{1}{45.5}{\nlx \textcolor{#3}{0.00}}
\PutLabel{16.2}{45.5}{\nlx \textcolor{#3}{0.15}}
\PutLabel{31.4}{45.5}{\nlx \textcolor{#3}{0.35}}
\PutLabel{46.6}{45.5}{\nlx \textcolor{#3}{0.70}}
\PutLabel{61.8}{45.5}{\nlx \textcolor{#3}{0.90}}
\PutLabel{77}{45.5}{\nlx \textcolor{#3}{1.30}}
\PutLabel{92.2}{45.5}{\nlx \textcolor{#3}{1.60}}
\PutLabel{107.4}{45.5}{\nlx \textcolor{#3}{2.00}}
\PutLabel{122.6}{45.5}{\nlx \textcolor{#3}{2.50}}
\PutLabel{137.8}{45.5}{\nlx \textcolor{#3}{3.00}}
\PutLabel{153}{45.5}{\nlx \textcolor{#3}{4.00}}
\PutLabel{168.2}{45.5}{\nlx \textcolor{#3}{5.00}}
\put(0,16.5){\includegraphics[height=1.6cm]{./simimages/PNG_#1/#1_EuclidYH_mz0.0.png}}
\put(15.2,16.5){\includegraphics[height=1.6cm]{./simimages/PNG_#1/#1_EuclidYH_mz0.15.png}}
\put(30.4,16.5){\includegraphics[height=1.6cm]{./simimages/PNG_#1/#1_EuclidYH_mz0.35.png}}
\put(45.6,16.5){\includegraphics[height=1.6cm]{./simimages/PNG_#1/#1_EuclidYH_mz0.5.png}}
\put(60.8,16.5){\includegraphics[height=1.6cm]{./simimages/PNG_#1/#1_EuclidYH_mz0.7.png}}
\put(76,16.5){\includegraphics[height=1.6cm]{./simimages/PNG_#1/#1_EuclidYH_mz0.9.png}}
\put(91.2,16.5){\includegraphics[height=1.6cm]{./simimages/PNG_#1/#1_EuclidYH_mz1.25.png}}
\put(106.4,16.5){\includegraphics[height=1.6cm]{./simimages/PNG_#1/#1_EuclidYH_mz1.5.png}}
\put(121.6,16.5){\includegraphics[height=1.6cm]{./simimages/PNG_#1/#1_EuclidYH_mz2.15.png}}
\put(136.8,16.5){\includegraphics[height=1.6cm]{./simimages/PNG_#1/#1_EuclidYH_mz3.0.png}}
\put(152,16.5){\includegraphics[height=1.6cm]{./simimages/PNG_#1/#1_EuclidYH_mz4.0.png}}
\put(167.2,16.5){\includegraphics[height=1.6cm]{./simimages/PNG_#1/#1_EuclidYH_mz5.0.png}}
\PutLabel{1}{29}{\nlx \textcolor{#3}{0.00}}
\PutLabel{16.2}{29}{\nlx \textcolor{#3}{0.15}}
\PutLabel{31.4}{29}{\nlx \textcolor{#3}{0.35}}
\PutLabel{46.6}{29}{\nlx \textcolor{#3}{0.50}}
\PutLabel{61.8}{29}{\nlx \textcolor{#3}{0.70}}
\PutLabel{77}{29}{\nlx \textcolor{#3}{0.90}}
\PutLabel{92.2}{29}{\nlx \textcolor{#3}{1.25}}
\PutLabel{107.4}{29}{\nlx \textcolor{#3}{1.50}}
\PutLabel{122.6}{29}{\nlx \textcolor{#3}{2.15}}
\PutLabel{137.8}{29}{\nlx \textcolor{#3}{3.00}}
\PutLabel{153}{29}{\nlx \textcolor{#3}{4.00}}
\PutLabel{168.2}{29}{\nlx \textcolor{#3}{5.00}}
\put(0,0){\includegraphics[height=1.6cm]{./simimages/PNG_#1/#1_VI_mz0.0.png}}
\put(15.2,0){\includegraphics[height=1.6cm]{./simimages/PNG_#1/#1_VI_mz0.05.png}}
\put(30.4,0){\includegraphics[height=1.6cm]{./simimages/PNG_#1/#1_VI_mz0.3.png}}
\put(45.6,0){\includegraphics[height=1.6cm]{./simimages/PNG_#1/#1_VI_mz0.4.png}}
\put(60.8,0){\includegraphics[height=1.6cm]{./simimages/PNG_#1/#1_VI_mz0.7.png}}
\put(76,0){\includegraphics[height=1.6cm]{./simimages/PNG_#1/#1_VI_mz0.9.png}}
\put(91.2,0){\includegraphics[height=1.6cm]{./simimages/PNG_#1/#1_VI_mz1.0.png}}
\put(106.4,0){\includegraphics[height=1.6cm]{./simimages/PNG_#1/#1_VI_mz1.3.png}}
\put(121.6,0){\includegraphics[height=1.6cm]{./simimages/PNG_#1/#1_VI_mz1.6.png}}
\put(136.8,0){\includegraphics[height=1.6cm]{./simimages/PNG_#1/#1_VI_mz2.0.png}}
\put(152,0){\includegraphics[height=1.6cm]{./simimages/PNG_#1/#1_VI_mz2.3.png}}
\put(167.2,0){\includegraphics[height=1.6cm]{./simimages/PNG_#1/#1_VI_mz2.7.png}}
\PutLabel{1}{12.5}{\nlx \textcolor{#3}{0.00}}
\PutLabel{16.2}{12.5}{\nlx \textcolor{#3}{0.05}}
\PutLabel{31.4}{12.5}{\nlx \textcolor{#3}{0.30}}
\PutLabel{46.6}{12.5}{\nlx \textcolor{#3}{0.40}}
\PutLabel{61.8}{12.5}{\nlx \textcolor{#3}{0.70}}
\PutLabel{77}{12.5}{\nlx \textcolor{#3}{0.90}}
\PutLabel{92.2}{12.5}{\nlx \textcolor{#3}{1.00}}
\PutLabel{107.4}{12.5}{\nlx \textcolor{#3}{1.30}}
\PutLabel{122.6}{12.5}{\nlx \textcolor{#3}{1.60}}
\PutLabel{137.8}{12.5}{\nlx \textcolor{#3}{2.00}}
\PutLabel{153}{12.5}{\nlx \textcolor{#3}{2.30}}
\PutLabel{168.2}{12.5}{\nlx \textcolor{#3}{2.70}}
\put(-3,6){\rotatebox{90}{\olx\textcolor{black}V-I}}
\put(-3,18){\rotatebox{90}{\olx\textcolor{black}Euclid Y-H}}
\put(-3,35){\rotatebox{90}{\olx\textcolor{black}F090-F150}}
\put(-3,50.5){\rotatebox{90}{\olx\textcolor{black}{$\mu$\arcmin} #8}}
\end{picture}
\caption{#2}
\label{#7}
\end{figure*}
\end{center}
}

\begin{document} 
\title{On the challenge of interpreting the morphology and color maps of high-{\it z} starburst galaxies
with the JWST and Euclid}
\titlerunning{On the challenge of interpreting color maps of of high-{\it z} starburst galaxies}
\authorrunning{Papaderos \& \"Ostlin}
   \author{
          Polychronis Papaderos      
          \inst{\ref{IA-CAUP},\ref{SU}}
          \and          
          G\"oran \"Ostlin
          \inst{\ref{SU}}
          }
\institute{
Instituto de Astrof\'{i}sica e Ci\^{e}ncias do Espaço - Centro de Astrof\'isica da Universidade do Porto, Rua das Estrelas, 4150-762 Porto, Portugal \label{IA-CAUP}
\and
Department of Astronomy, Oskar Klein Centre; Stockholm University; SE-106 91 Stockholm, Sweden \label{SU}\\
             \email{papaderos@astro.up.pt, ostlin@astro.su.se}
             }
\date{Received ?; accepted ?}
\abstract{Morphology and color patterns hold fundamental insights into the early assembly history and physical ingredients
(e.g., stellar and nebular emission, dust obscuration) of high-\zet\ galaxies rapidly building up their stellar mass at a
high specific star formation rate (sSFR).
However, a 2D reconstruction of rest-frame color maps of such systems from multi-band imaging data is a non-trivial task.
This is mainly because the spectral energy distribution (SED) of high-sSFR galaxies near and far is spatially inhomogeneous
and thus the common practice of applying a spatially constant "morphological" \kc-correction tailored to their integral (luminosity-weighted) SED almost unavoidably leads to serious and systematic observational biases.
In this study we use the nearby blue compact galaxy \object{Haro 11} to illustrate how the spatial inhomogeneity of the SED
in the visual and rest-frame UV impacts the morphology and color maps in the observer's frame (ObsF) visual and NIR, and potentially
affects the physical characterization of distant starburst galaxies with the \emph{James Webb} Space Telescope (JWST) and Euclid.
Based on MUSE integral field spectroscopy and spectral modeling we first examine the elements shaping the spatially varying optical
SED of Haro~11, namely intrinsic stellar age gradients, strong nebular emission (\ne) and its spatial decoupling from the ionizing
stellar background, and differing extinction patterns in the stellar and nebular component both spatially and in their amount.
Our simulations show, inter alia, that an optically bright yet dusty star-forming (SF)
region may evade detection whereas a gas-evacuated (thus, potentially Lyman continuum photon-leaking) region with weaker SF activity
can dominate the ObsF (rest-frame UV) morphology of a high-$z$ galaxy.
It is also demonstrated that ObsF color maps are drastically affected by the spatial inhomogeneity of the SED, being especially prone to strong emission lines moving in and out of filter passbands depending on \zet, and, if taken at face value, leading to erroneous conclusions about the nature, evolutionary status and dust content of a galaxy. A significant additional problem in this regard stems from the uncertain prominence of the 2175 \AA\ extinction bump that translates to appreciable ($\sim$0.3 mag) inherent uncertainties in rest-frame color maps of high-\zet\ galaxies.}

\maketitle
\section{Introduction \label{intro}}
Recently, \citet[][hereafter \tref{P23}]{P23} drew attention to the chromatic surface brightness modulation (\cmod) effect and to the serious biases that its neglect introduces in studies of higher-redshift (\zet$>$0.1) galaxies. \cmod\ primarily results from the differential surface brightness ($\mu$) dimming of UV-faint (non-star-forming or dusty) regions and the simultaneous $\mu$-enhancement of UV-bright (star-forming) regions in a higher-\zet\ galaxy whose rest-frame UV is recorded in the optical and near-infrared (NIR) ObsF.
In the case of extreme emission-line galaxies (EELGs), such as nearby blue compact galaxies (BCGs) and green peas (GPs), as well as high-\zet\ protogalaxies rapidly assembling their stellar mass \mstar\ at a high sSFR, \cmod\ is further amplified by strong \ne\ lines that selectively enhance $\mu$ depending on filter, redshift and rest-frame emission-line equivalent width (EW) patterns \citep[][hereafter \tref{PO12}]{PO12}. SF feedback in these systems leads to extended, approximately exponential \ne\ halos with a high ($>10^3$ \AA) and frequently outwardly increasing EW (\tref{P02}).
Thus, intense and spatially inhomogeneous \ne\ adds a significant level of complexity to \cmod, further contributing to the possible amplifation, erase or inversion of ObsF radial color gradients in a high-\zet\ galaxy (\tref{P02,P23}).

The previously overlooked \cmod\ effect also strongly impacts the ObsF morphology and structural properties (e.g., bulge-to-disk ratio, S\'ersic exponent $\eta$, effective radius \reff) of higher-\zet\ galaxies. Its correction is therefore fundamental to the objective characterization of the galaxy assembly history and the variation across \zet\ of galaxy scaling relations (e.g., the mass-metallicity, Tully-Fisher, and bulge vs. super-massive black hole relation).
However, this task (the 2D reconstruction of rest-frame from ObsF photometric properties) is non-trivial, as it requires a spatially resolved \kc-correction that is adapted to the evolving 2D rest-frame SED of a galaxy. This, in turn, depends on several interlinked factors, such as the star formation history (SFH) and dynamical make-up (e.g., role of mergers and stellar migration) of a galaxy, and its chemical properties and \ne- and dust attenuation patterns.

Since \tref{P23} have mainly focussed of the implications of \cmod\ for our understanding of bulge growth in spiral galaxies, touching only peripherically upon EELGs, we here supplement our previous considerations by a concise empirical inspection of \cmod\ effects in the latter class of objects.
To this end, we use as an illustrative example the nearby ($D=82$ Mpc) luminous BCG \object{Haro 11} \citep{BO86}. This dwarf galaxy merger \citep{O15} is a bright \lya\ emitter \citep{Hayes07,O09} and the nearest confirmed Lyman continuum (\lyc) leaking galaxy \citep{Bergvall06,Leitet11}, properties that have made it a popular local analog of galaxies in the epoch of reionization. It contains three high-$\mu$ knots (A--C) with quite different properties \citep{O21,Sirressi22} in terms of emission line strength, ionization level and reddening, and therefore serves as a good example of the intrinsic complexity of a starburst galaxy, and how it is amplified through \cmod.

Section~\ref{IFS} provides a brief summary of the physical characteristics of \object{Haro 11} as obtained from integral field spectroscopy (IFS) and spectral synthesis, and is limited to a few aspects with special relevance to \cmod, in particular, the spatially inhomogeneous SED of the BCG as the result of its intrinsic stellar age gradients, a large-scale spatial decoupling of \ne\ from ionizing young stellar clusters (YSCs), and differing extinction patterns in the stellar and nebular component both in their amount and spatially.
In turn (Sect.~\ref{zSim}) we use the method introduced in \tref{P23} to compute out to \zet=5.4 the effect of \cmod\ on the morphology and color maps of \object{Haro 11} in \emph{James Webb} Space Telescope (JWST) and Euclid filters. Our conclusions are summarized in Sect.~\ref{sum}, and the appendix (Sect.~\ref{app}) provides supplementary notes on the processing of IFS data for \object{Haro 11} as well as simulations of the \cmod\ effect for a representative set of galaxy morphologies
(also available at this {\color{red}\tt hyperlink:tbd}).
\begin{center}
\begin{figure*}
\begin{picture}(200,130)
\put(-1,73){\includegraphics[height=5.8cm]{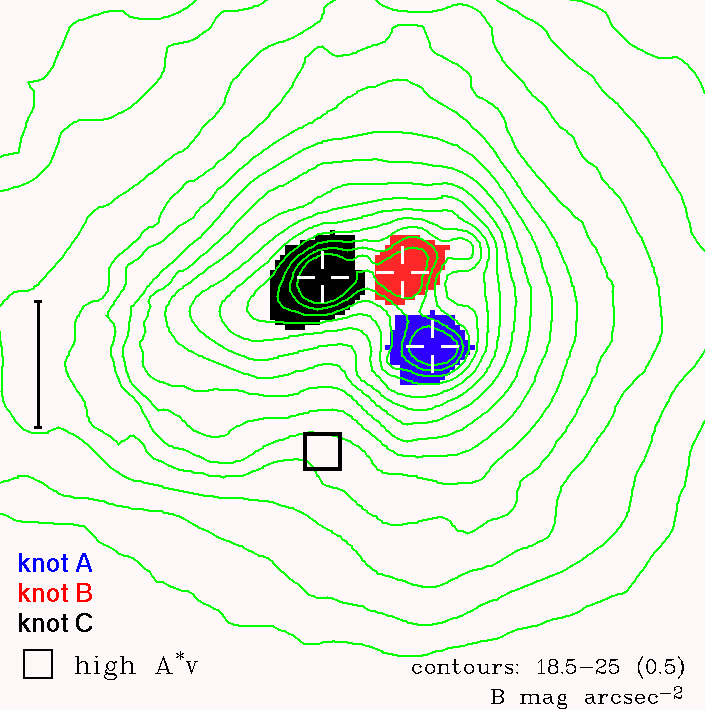}}
\put(62.4,73){\includegraphics[height=5.8cm]{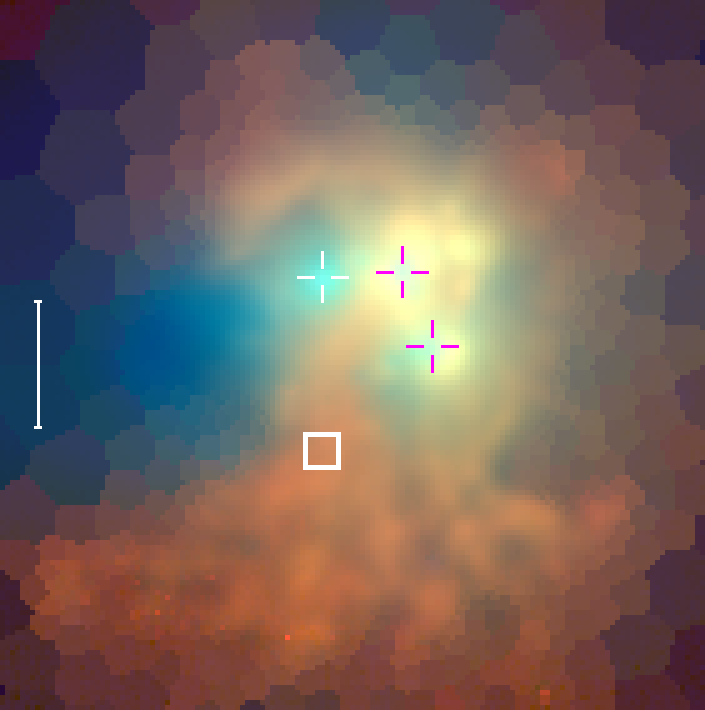}}
\put(126.5,73){\includegraphics[height=5.8cm]{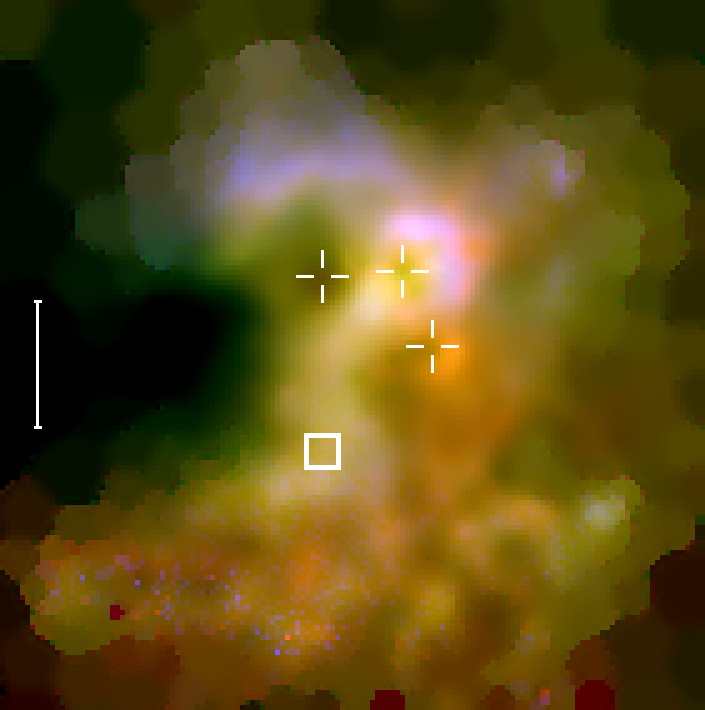}}
\PutLabel{0}{126}{\mfont \textcolor{black}{a}}
\PutLabel{116.5}{126}{\mfont \textcolor{white}{b}}
\PutLabel{180.5}{126}{\mfont \textcolor{white}{c}}
\put(-2,42){\includegraphics[height=2.9cm]{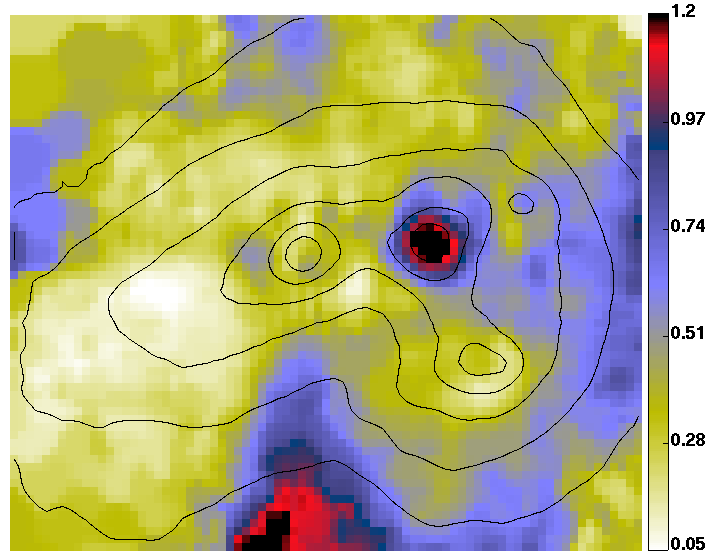}}
\put(36,42){\includegraphics[height=2.9cm]{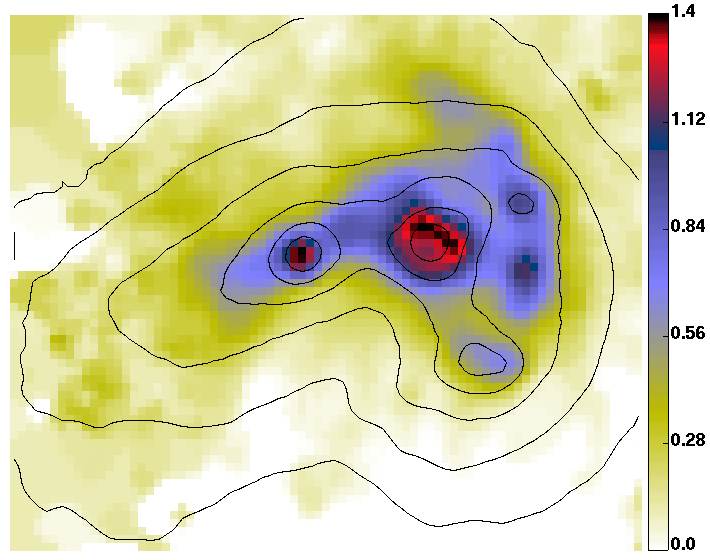}}
\put(74,42){\includegraphics[height=2.9cm]{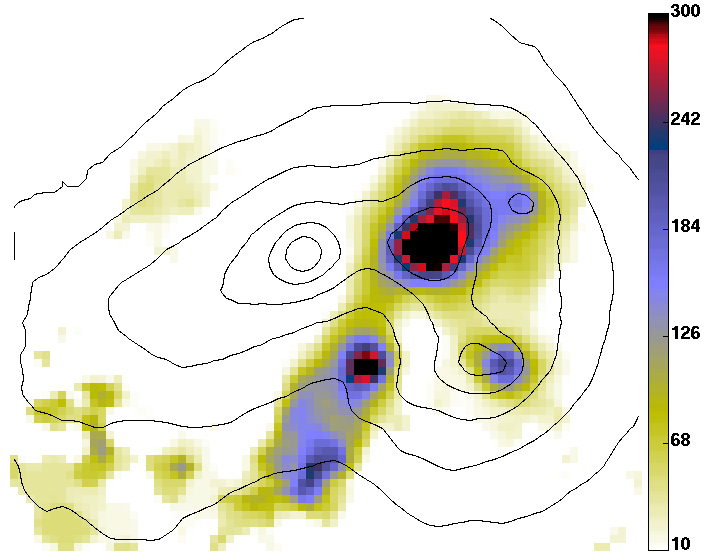}}
\put(112,42){\includegraphics[height=2.9cm]{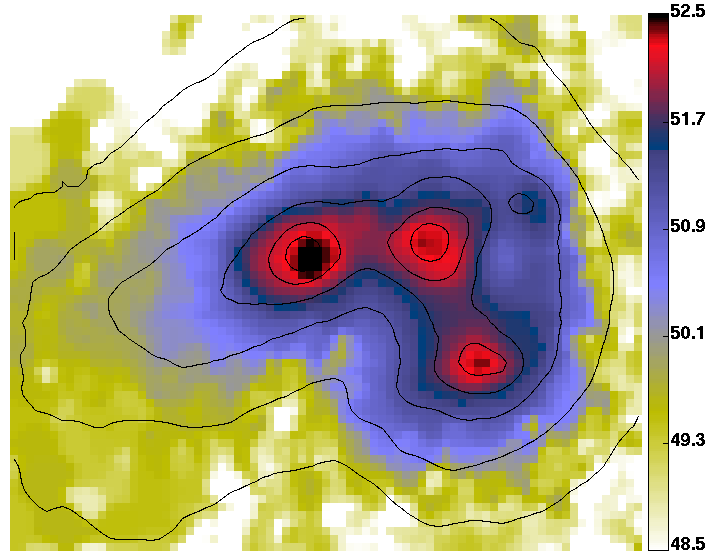}}
\put(150,42){\includegraphics[height=3cm]{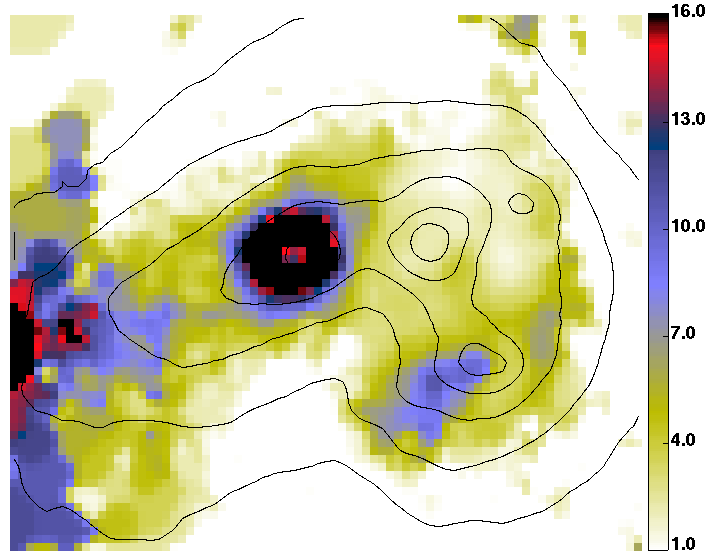}}
\PutLabel{-0.5}{66}{\mfont \textcolor{black}{d}}
\PutLabel{38}{66}{\mfont \textcolor{black}{e}}
\PutLabel{76}{66}{\mfont \textcolor{black}{f}}
\PutLabel{114}{66.5}{\mfont \textcolor{black}{g}}
\PutLabel{152}{66.5}{\mfont \textcolor{black}{h}}
\put(-1,40){\includegraphics[width=4cm, angle=-90]{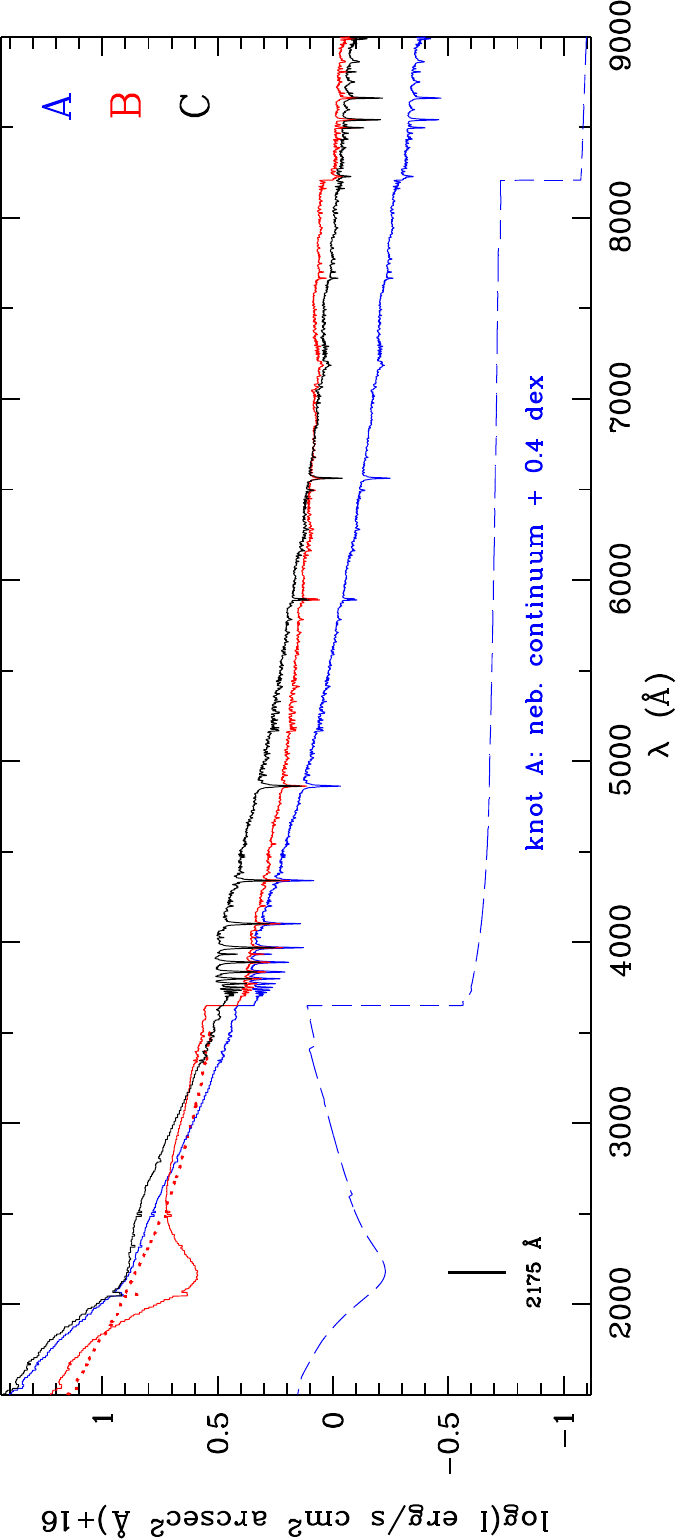}}
\put(96,40){\includegraphics[width=4cm, angle=-90]{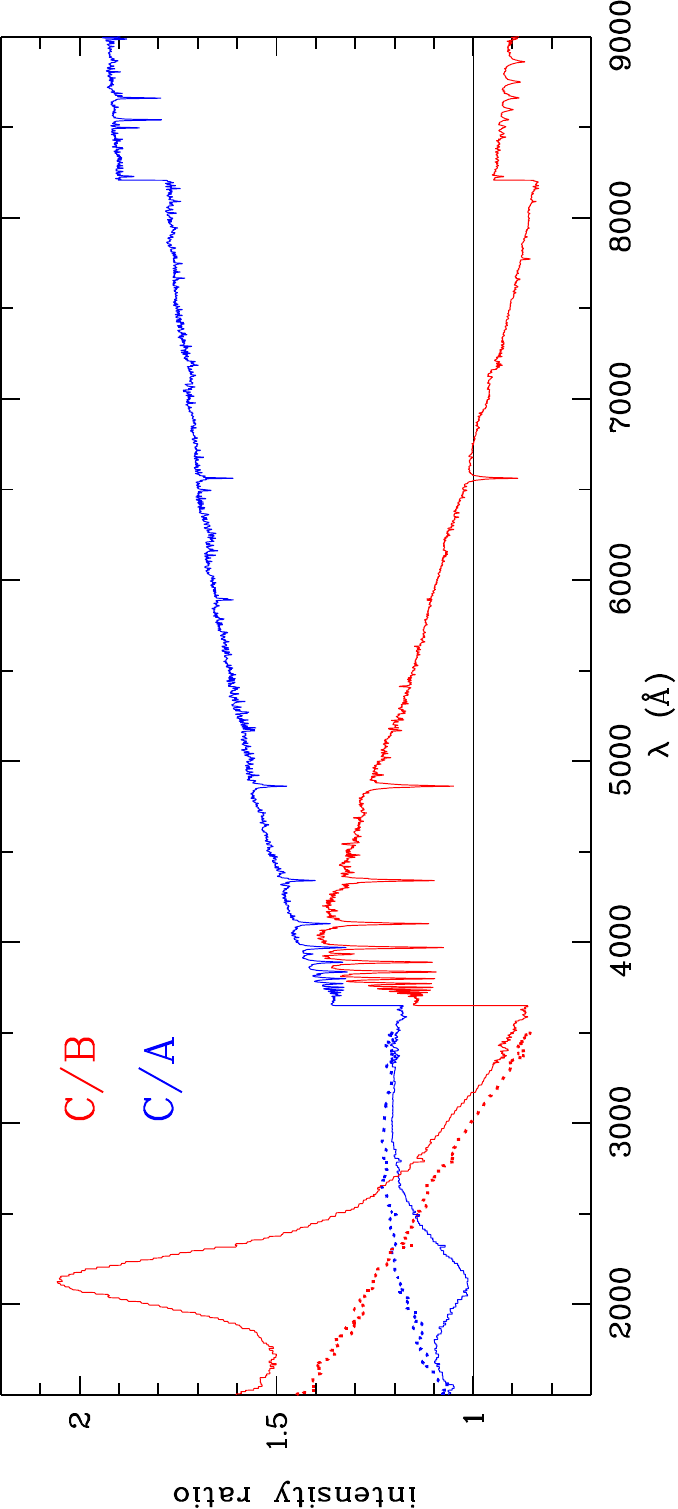}}
\PutLabel{90}{36}{\mfont \textcolor{black}{i}}
\PutLabel{185}{36}{\mfont \textcolor{black}{j}}
\PutLabel{59}{58}{\mfont \textcolor{black}{k}}
\PutLabel{119}{58}{\mfont \textcolor{black}{l}}
\end{picture}
\caption{Synopsis of some key characteristics of \object{Haro 11} ($D$=82 Mpc). \brem{a:} $B$-band contour map computed from MUSE IFS data (cf. \tref{P23}). Crosses mark the intensity maxima of regions A--C. North is up and east to the left. The square marks the \exstars\ maximum in the southerm half of the BCG. The vertical bar corresponds to a linear scale of 2 kpc.
\brem{b}: True-color composite of \ewha, [O{\sc iii}]$_{5007}$ flux and the \ne-free stellar continuum between 6390 \AA\ and 6490 \AA\ (red, green and blue, respectively)
and \brem{c:} of the \ha, [N{\sc ii}]$_{6583}$, and [O{\sc i}]$_{6300}$ EW.
Regions with a shock-enhanced [O{\sc i}]/\ha\ ratio appear blueish and those with an elevated [N{\sc ii}]/\ha\ ratio greenish.
\brem{d:} Stellar extinction \exstars\ (mag). Contours in F115W (AB system) go from 16.7 to 21.7 \sbb\ in steps of 1 mag.
\brem{e:} Nebular extinction \exneb\ (mag) computed from the \ha/\hb\ ratio.
\brem{f:} Electron density (cm$^{-3}$).
\brem{g:} Logarithm of the intrinsic (non-attenuated) \lyc\ production rate \NLyC\ (s$^{-1}$).
\brem{h:} $\tau$ ratio. Dark areas depict regions with a high local \lyc\ photon leakage.
\brem{i:} UV-through-optical SED (in $10^{-16}$ erg/(s cm$^{2}$ \AA\ $\sq\arcsec$)) of the total (stellar and nebular) continuum in knots A--C,
attenuated following \citet[][hereafter CCM]{Cardelli89}. The dashed curve shows the nebular continuum SED of knot A.
\brem{j:} SED ratio C/A and C/B.
The dotted curve between 1500 and 3500 \AA\ illustrates the effect of attenuation following \citet[][hereafter CAL]{Calzetti00} on the stellar SED of knot B (panel \brem{i}) and on the C/A and C/B intensity ratio (panel \brem{j}).
\label{fig1}}
\end{figure*}
\end{center}

\vspace*{-1cm}
\section{The spatially inhomogeneous SED of Haro 11 \label{IFS}}
The analysis of \object{Haro 11} is based on IFS data acquired with MUSE \citep{Bacon14} and reduced as described in \citet{Menacho19}. 
Following \tref{P23}, the IFS data cube was modeled with the population synthesis code \SL\ \citep{Cid05}, both in its original form and
after subtraction of the nebular continuum, and, as a consistency check, also with \FD\ \citep{GP17}.
In turn (Sect.~\ref{zSim}), the 2D UV-through-IR stellar SED of the BCG was computed from the best-fitting population vectors,
and after addition of \ne, used to simulate the morphology and color patterns of the galaxy out to \zet=5.4.

Figure~\ref{fig1} illustrates the spatial inhomogeneity of the SED within and around the three SF knots (A-C; panel \brem{a}) of \object{Haro 11},
and the implications it has for the rest-frame UV morphology of the BCG.
A salient feature in panel \brem{b} is the spatial decoupling of nebular and stellar emission, as a typical feature of starburst galaxies \citep[\tref{P02},][hereafter \tref{O03}]{Ostlin03}. \ne\ is especially pronounced in the N-S direction, protruding $\sim$2 kpc SE from knot B. 
Contrary to regions A and B where \ne\ is strong, with the \ewha\ reaching average values of $\sim$500 \AA\ and $>$900 \AA, respectively (cf. Fig.~\ref{app:Haro11}), it is comparatively weak \ne\ in region C (160 \AA, decreasing to $<$40 \AA\ at its eastern half), which suggests that gas there is partially blown away and SF is about to be extinguished. A hint that this begun only recently ($\la$40 Myr ago), as suggested by \citet{Sirressi22},
comes from the shock-enhanced [O{\sc i}]/\ha\ ratio all over the extended northern rim (panel \brem{c}) witnessing the still ongoing release of mechanical luminosity by SNe.

As apparent from panels \brem{d} and \brem{e}, stellar and nebular emission (\exstars\ and \exneb, respectively) display substantially different extinction patterns: the average \exstars\ in knot B ranges between 0.5 and $\sim$1~mag (\SL\ and \FD, respectively), peaking centrally, whereas it is modest ($\sim$0.3 mag) in regions A\&C. This is consistent with HST imaging revealing a dust lane in B. A hard ($>$3 keV) X-ray source detected there hints at a low-luminosity AGN powered by an intermediate-mass black hole of $<5 \times 10^7$ \msun\ \citep{Prestwich15,Gross21}.
Accretion-powered activity is further suggested by the classification of region B as Composite on the basis of the [O{\sc iii}]/\hb\ vs. [N{\sc ii}]/\ha\ ratio, its relatively high electron density $n_{\rm e}$ ($\sim 300$ cm$^{-3}$, and locally up to 440 cm$^{-3}$; panel \brem{f}), and its non-thermal radio emission \citep{LeReste23}.

As for \exneb, it is highest in region B (up to 1.5 mag) 
and modest in C and A (0.8 and 0.46 mag, respectively). Of special importance to \cmod\ effects is the partial spatial decoupling of \exneb\ and \exstars\ both with respect to each other and to the stellar surface density \mustar: \exneb\ shows several local maxima west of region B, whereas \exstars\ forms a contiguous rim with $\sim$0.6 mag all over the SW periphery of the galaxy and reaching $\ga$1~mag $\sim 7$\arcsec\ south of its knot C.
Interestingly, this region (marked with a square in panels \brem{a}-\brem{c}) shows double velocity components \citep[interpreted as the overlap of the two disks participating in the merger,][]{O15,Menacho21}, and is nearly cospatial with the southern \ewha\ feature where a chain of higher-$n_{\rm e}$ pockets might suggest denser material entrained in a starburst-driven outflow (panel \brem{f}).
\exneb\ here is low ($<$0.15 mag), so the strongly reddened stellar component likely lies on the far side of the nebular gas.

Spectral modeling suggests that the region with the highest mass fraction of stars younger than 100 Myr (14\%) is C,
followed by A (9\%) and B (3\%). It should be noted, however, that the SFH of region B is somehow uncertain,
not only because the 3D geometry of dust probably departs from a simple foreground screen model (as already indicated by the dust lane),
but also due to a possible contribution of a non-thermal AGN continuum that could appreciably bias spectral modeling studies \citep{Cardoso16}.
Nevertheless, taking the inferred SFH (Fig.~\ref{app:SFHs}) at face value, region B is dominated by old ($>$3 Gyr) stars with only a tiny substrate of young stars overshining the old stellar background, whereas region C has experienced prolonged SF over several Gyr and still contains an appreciable mass fraction of ionizing stars.
As for region A, SF appears to ramp up since $\sim$100 Myr, roughly the time indicated by numerical simulations for the first crossing of the two galaxies that participated in the \object{Haro 11} merger (Ejdetj\"arn et al. 2023, in prep.).

The morphology and color patterns of \object{Haro 11} result from i) the spatially inhomogeneous SED of its stellar populations,
ii) the (SFH- and metallicity-dependent) intensity $\mu_{\rm Ly{\sc c}}$ (\lbb) of \lyc\ radiation produced and
iii) its reprocessing into \ne\ both on the spot and out to kpc away from it,
and iv) the spatially differing extinction patterns of stars and \ne.

Evidently, in view of these multiple factors, a meaningful reconstruction of the 2D rest-frame UV (ObsF optical and NIR) characteristics of a distant analog of \object{Haro 11} is barely possible with the usual assumption of a spatially constant "morphological k-correction" adapted to the integral (thus, luminosity-weighted)
SED of the BCG (see discussion in \tref{PO12} and \tref{P23}).
One reason for this is that an optically bright yet dust- and metal-rich SF region might turn out to be fainter in the restframe UV than a dust-poor region with a lower SFR.
This is exemplified in panels \brem{i}\&\brem{j}: the stellar SED of the optically bright SF region B is in the UV fainter than that of the lower-SFR region C.
Although of comparable monochromatic luminosity at $\lambda$(\ha)), the SEDs of these regions diverge in the UV where C becomes at $\sim 2175$ \AA\ by a factor $\sim$2 ($\sim$1.3) brighter than B when CCM (CAL) is assumed.
In the case of a high-\zet\ analog of \object{Haro 11}, this translates into a strong discrepancy between ObsF and rest-frame morphology, especially at redshifts
where the broad region around the 2175 \AA\ absorption bump (1850--2500 \AA) falls within a filter transmission curve.
This happens, for example, at \zet=1.5 (4.3) in the $V$ (JWST F115W) filter.

Another determinant of the morphology is the temporal and spatial characteristics of the production of \lyc\ radiation and its reprocessing into \ne\ within the evolving multi-phase gas topology of a starburst galaxy.
For a given initial mass function and stellar binarity, the intrinsic \lyc\ production rate (\NLyC) depends on the
SFH and stellar metallicity. It can be seen from panel \brem{g} that region C, despite its faint \ne, dominates the \lyc\ photon budget, reaching centrally a value log($\mu_{\rm LyC}$ \lbb)$\sim$53.7, slightly above that in region B (53.5) and A (53.4),
in qualitative agreement with Fig.~9 in \citet{Hayes07}. Region C is also the strongest in \lya\ \citep{O21}.

Important in this context is also that the fraction of \lyc\ photons reprocessed into \ne\ in situ relative to those doing so on larger spatial scales (or perhaps even leaking out into the circumgalactic space) shows large spatial variations. In \object{Haro 11} this is reflected in the spatial decoupling of \ne\ from the ionizing YSCs and the spatial anti-correlation of \msb\ and \ewha\ this process typically leads to (\tref{P02},\tref{O03},\tref{P23}).
An estimate of the local \lyc\ escape fraction can be obtained with a variant of the $\tau$ ratio, which was defined in \citet[][hereafter \tref{P13}]{P13} as the
ratio of the H$\alpha$ luminosity expected under standard conditions ($n_{\rm e}=100$ cm$^{-3}$ and an electron temperature $T_{\rm e} = 10^4$ K) from the \lyc\ output from post-AGB sources to the observed value.
Here, in computing $\tau$ we use instead the total ionizing output from stars after attenuation by \exstars\ and assuming CCM.
A $\tau=1$ corresponds to an equilibrium state where \lyc\ production exactly balances the observed \ha,
whereas larger values translate to a \lyc\ leakage fraction of $1 - (1/\tau)$.
The $\tau$ in regions A and B ($\sim$2) is within uncertainties (cf. \tref{P13}) consistent with in-situ reprocessing of the total \lyc\
output from stars along the line of sight, whereas that in region C (16--20) implies that the bulk of \lyc\ radiation locally escapes, and is
eventually captured at larger radii by denser neutral gas mixed within the extended nebular halo.

Finally, another factor shaping the ObsF morphology of a higher-\zet\ analog of \object{Haro 11} is the attenuation of the UV SED in individual sightlines. Substantial uncertainties in this regard stem from the fact that \exstars\ and \exneb\ can differ both spatially and in their amount (Fig.~\ref{fig1}d\&e), and because it is unclear whether standard parameterizations such as CAR or CAL are valid both for stellar and nebular emission on subgalactic scales and across \zet. As pointed out above, specially relevant in this regard is the prominence of the 2175 \AA\ bump.
Even though there is no support in favor of or against CCM in \object{Haro 11} specifically,
the existing evidence for the 2175 \AA\ feature showing large variations from galaxy to galaxy \citep[e.g.,][see also discussion in \tref{Reddy et al. 2018}]{Narayanan18} entails significant uncertainties for studies of distant galaxies, as apparent from Fig.~\ref{fig1}\brem{i}\&\brem{j} and the simulations of \cmod\ effects in Sect.~\ref{zSim}.
%
\section{A high-\zet\ analog of Haro 11\label{zSim}}
\begin{center}
\begin{figure}[t]
\begin{picture}(86,58)
\put(0,38){\includegraphics[height=1.9cm]{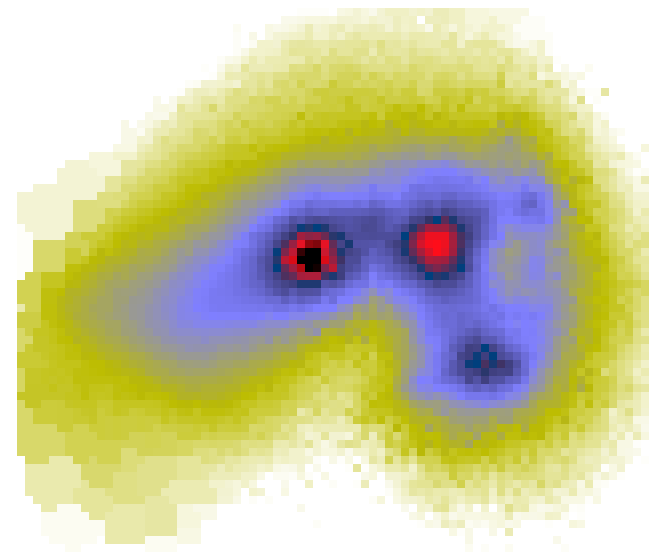}}
\put(22,38){\includegraphics[height=1.9cm]{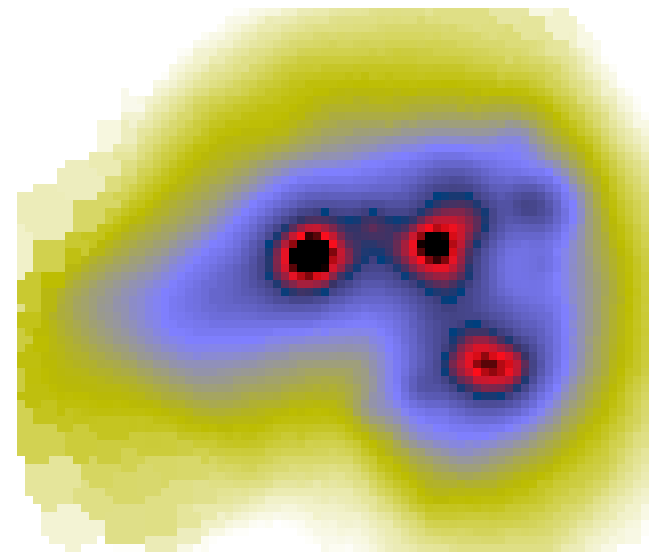}}
\put(44,38){\includegraphics[height=1.9cm]{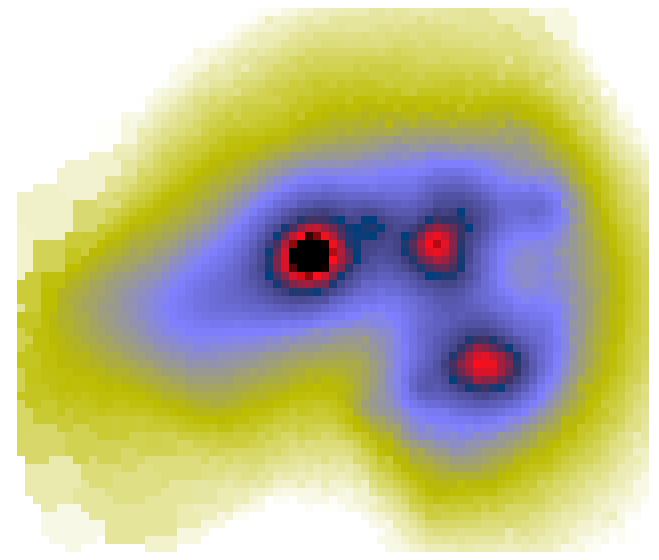}}
\put(67,38){\includegraphics[height=1.9cm]{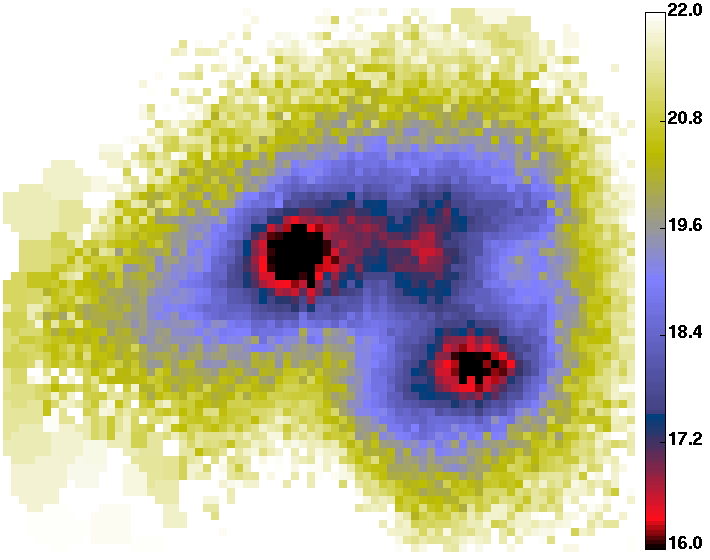}}
\put(0,0){\includegraphics[width=8.6cm]{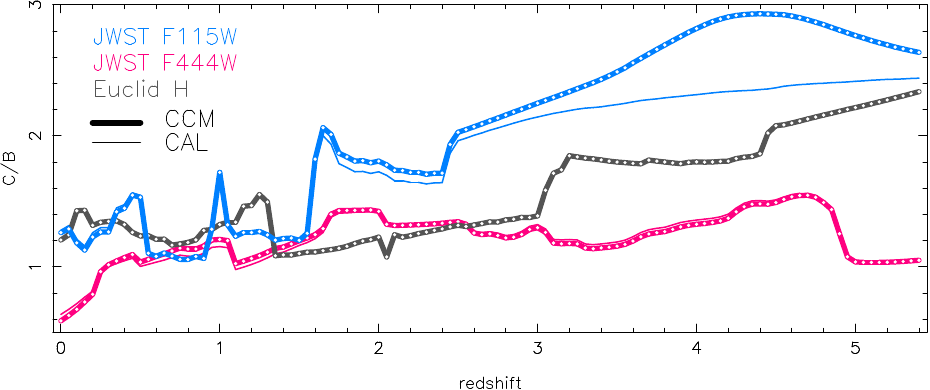}}
\PutLabel{2}{53.5}{\nlx \textcolor{black}{z=0}}
\PutLabel{24}{53.5}{\nlx \textcolor{black}{1.45}}
\PutLabel{47}{53.5}{\nlx \textcolor{black}{1.7}}
\PutLabel{70}{53.5}{\nlx \textcolor{black}{4.4}}
\end{picture}
\caption{Variation of the morphology of \object{Haro 11} vs. \zet.
\brem{top:} Simulated images at \zet=0, 1.45, 1.7 and 4.4 in the JWST filter F115W.
The maps show the \emph{reduced} surface brightness $\mu\arcmin$ (\sbb) computed taking only
bandpass shift and wavelength stretching into account, and refer to the CCM extinction law.
\brem{bottom:} Ratio C/B of the mean intensity in region C and B within their central 1\farcs6
in the filters F115W and F444W (JWST) and H (Euclid). Thick and thin curves refer, respectively, to CCM and CAL.
\label{dmag}}
\end{figure}
\end{center}

The discussion next is meant to illustrate the complexity that \cmod\ introduces in the spatially resolved reconstruction of the rest-frame properties (morphology, color maps) of high-\zet\ EELGs.
From the simulated images of \object{Haro 11} in the JWST filter F115W (Fig.~\ref{dmag}) it is apparent that the optically bright starburst
region B virtually disappears at \zet=4.4 despite its very high restframe \ewha.
The effect of \cmod\ on the ObsF morphology can better be quantified from the lower panel which compares the ratio C/B of the mean
intensity in region C and B in the filters F115W and F444W (JWST) and H (Euclid) as a function of \zet.
This ratio rises in the F115W filter from $\sim$1.2 at \zet=0 to $\sim$3 ($\sim$2.4) at \zet=4.4 when assuming CCM (CAL),
which implies that the less obscurred region C dominates at high \zet\ while knot B is barely visible.

\begin{center}
\begin{figure}
\begin{picture}(86,80)
\put(5,61.5){\includegraphics[width=8.2cm]{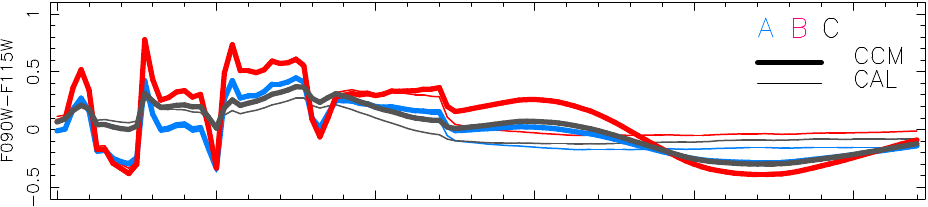}}
\put(5,42.4){\includegraphics[width=8.2cm]{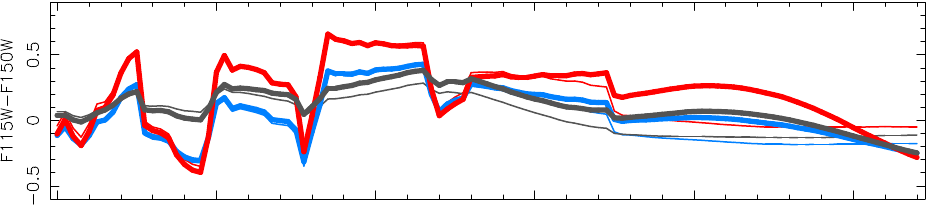}}
\put(5,22.4){\includegraphics[width=8.2cm]{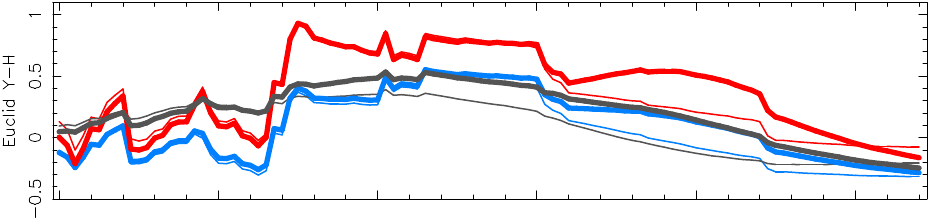}}
\put(5,0){\includegraphics[width=8.2cm]{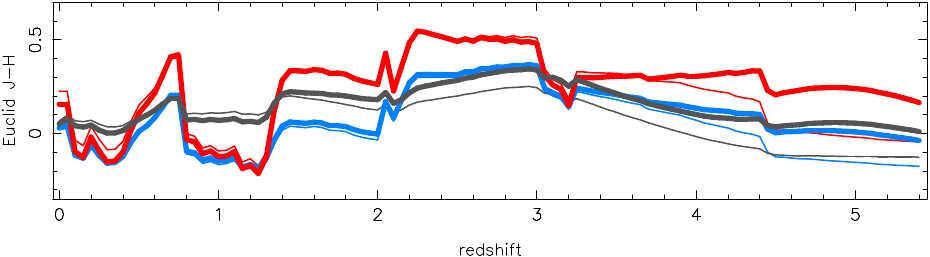}}
\end{picture}
\caption{Simulated ObsF color of regions A-C vs. \zet\ in the case of CCM and CAL
(see Fig.~\ref{app:col_vs_z} for an extended set of colors).
\label{col_vs_z}}
\end{figure}
\end{center}

Relevant to the morphological characterization of a high-\zet\ EELG analog of \object{Haro 11} is also the intrinsic extinction its restframe UV SED
experiences. In this regard, the 2175 \AA\ absorption bump, prominent in CCM whereas absent in CAL, introduces a substantial inherent uncertainty. One implication of this is that color maps combining filters with a substantially different central wavelength $\lambda_0$, with only one of those encompassing the redshifted 2175 \AA\ bump (e.g., F115W and F444W at \zet$\sim$4.5), can systematically be biased at a level of $\ga$0.3 mag (cf., e.g., Fig.~\ref{col_vs_z}). On the other hand, precisely this weakness, offers an avenue toward observationally constraining the strength of the 2175 \AA\ bump when the intrinsic UV SED of a high-\zet\ source is known.

Another salient feature in Fig.~\ref{dmag} is the local discontinuities of the C/B ratio at various \zet. These can be understood with the help
of Fig.~\ref{zetfil} as primarily the effect of strong emission lines reshifted in and out of filter transmission curves.
For instance, this is the case at \zet=1 (1.7) when \ha\ (H$\beta$ and [O{\sc iii}]$_{4959,5007}$) moves out of the F115W filter, or at \zet$\sim$2.5 when this happens for the [O{\sc ii}]$_{3727,29}$ doublet and the Balmer jump (BJ).
Same applies to Euclid $H$ at, e.g., \zet$\sim$3.2 (4.4) when the H$\beta$ and [O{\sc iii}] lines (the UV [O{\sc ii}] doublet and the BJ) drop out.
Important in the context of EELGs is that, largely because of the sharp boundaries of the JWST and Euclid NIR filter transmission curves,
even a small ($\sim$0.1) change in \zet\ can lead to a strong (up to $\sim$0.5 mag) enhancement or dimming (Fig.~\ref{app:mmag})
in one particular filter, hence to a reversion of colors (Figs.~\ref{col_vs_z} and \ref{app:col_vs_z}).

\begin{center}
\begin{figure}
\begin{picture}(86,46)
\put(0,24){\includegraphics[height=2.3cm]{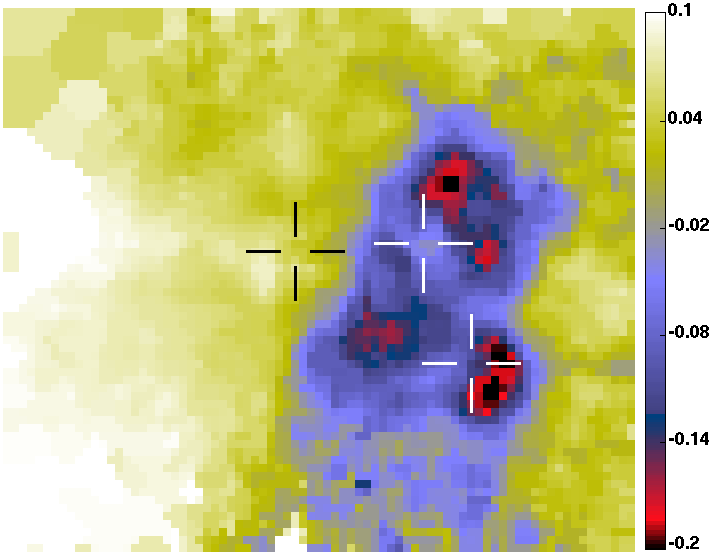}}
\put(30,24){\includegraphics[height=2.3cm]{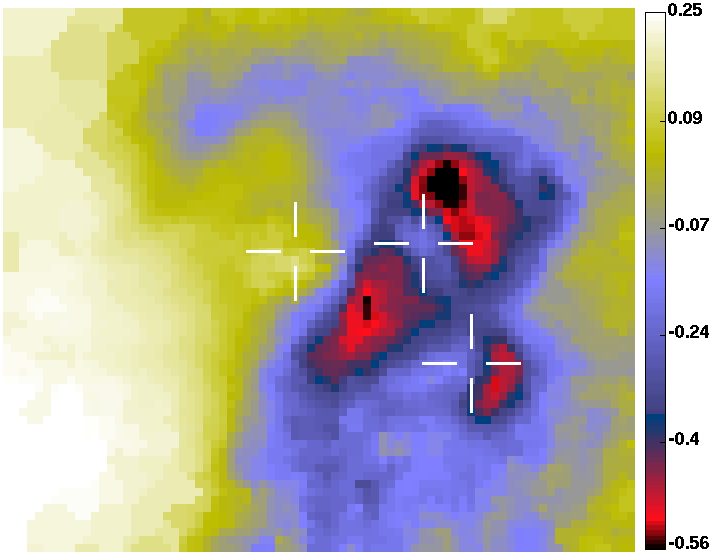}}
\put(60,24){\includegraphics[height=2.3cm]{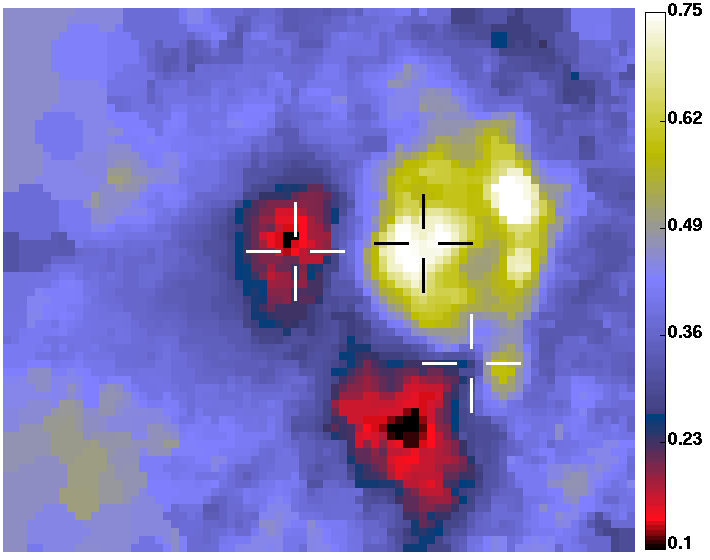}}
\PutLabel{1}{44}{\olx \textcolor{black}{F115W-F150W}}
\PutLabel{1}{25.5}{\nlx \textcolor{black}{z=0}}
\PutLabel{31}{25.5}{\nlx \textcolor{black}{0.9}}
\PutLabel{61}{25.5}{\nlx \textcolor{white}{1.7}}
\put(0,0){\includegraphics[height=2.3cm]{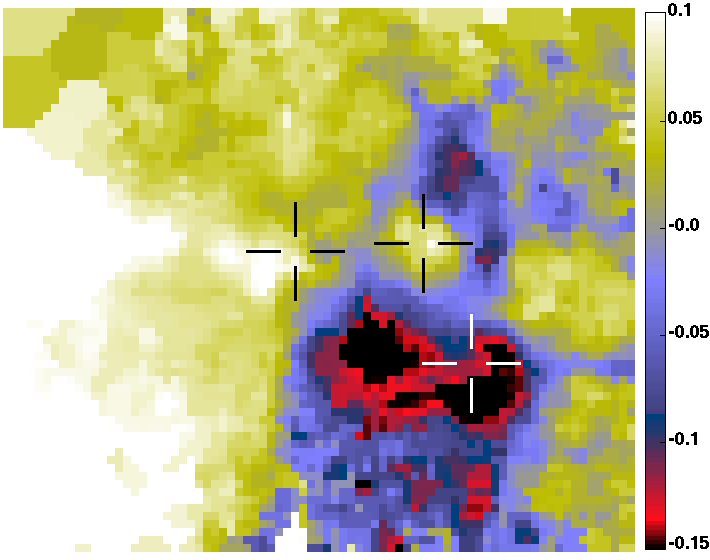}}
\put(30,0){\includegraphics[height=2.3cm]{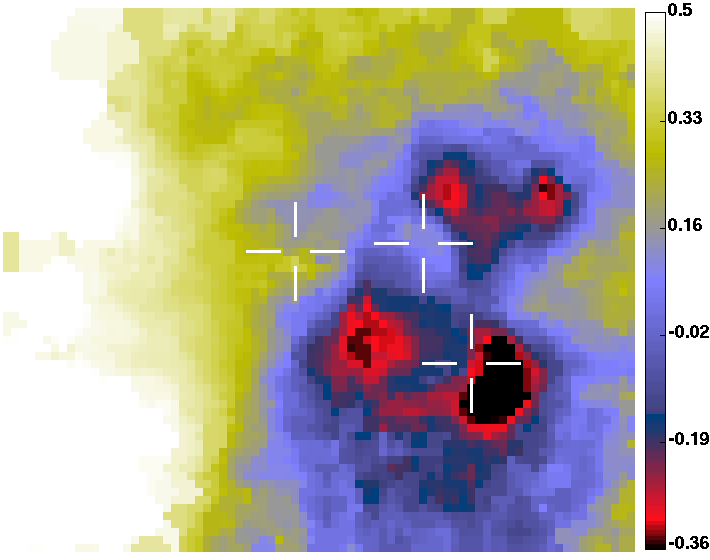}}
\put(60,0){\includegraphics[height=2.3cm]{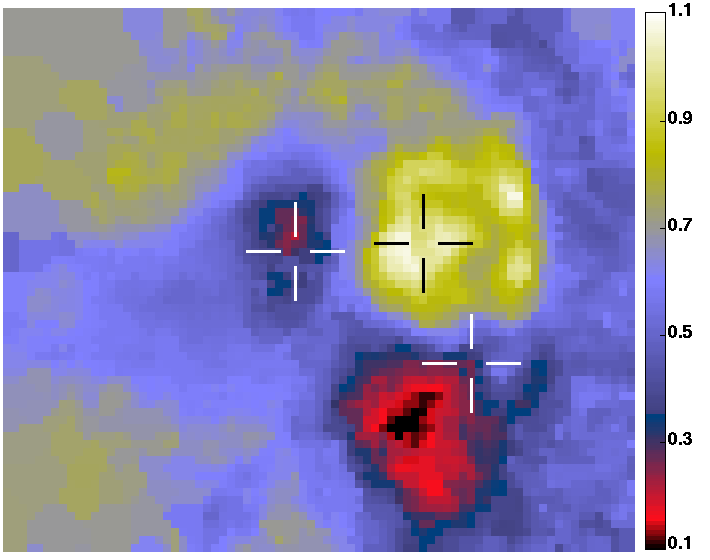}}
\PutLabel{1}{19.6}{\olx \textcolor{black}{Euclid Y-H}}
\PutLabel{1}{1.5}{\nlx \textcolor{black}{z=0}}
\PutLabel{31}{1.5}{\nlx \textcolor{black}{1.25}}
\PutLabel{61}{1.5}{\nlx \textcolor{white}{1.5}}
\end{picture}
\caption{Simulated color maps of \object{Haro 11} in JWST F115W--F150W at \zet=0, 0.9 and 1.7 (top)
and Euclid Y-H (bottom) at \zet=0, 1.25 and 1.5. Crosses mark the position of regions A-C
(see Fig.~\ref{app:Haro11} for supplementary material).
\label{zSim:colormaps}}
\end{figure}
\end{center}

\vspace*{-1cm}
Figure~\ref{zSim:colormaps} offers an example for how deceptive color maps of a distant starburst galaxy can be:
whereas region B is rather inconspicuous in F115W--F150W at \zet=0 and 0.9, it is red at \zet=1.7, just like at \zet=1.5
for the Euclid Y--H. The spatially inhomogeneous SED of an EELG naturally leads to complex color patterns that can strikingly
vary with \zet\ and which show little spatial correspondence to \msb\ as they are mostly dictated by emission-line EWs.
This together with the additional effect of \cmod\ on galaxy morphology (Fig.~\ref{dmag}; see also \tref{P23}) renders the
physical characterization of high-\zet\ EELGs (starburst galaxies, AGN) via imaging data a non-trivial endeavor.
As pointed out in \tref{PO12}, applying a standard (spatially constant) \kc-correction does not alleviate the problem but it
potentially aggravates it in a hardly predictable manner.

\vspace*{-0.4cm}

\vspace*{-1.2cm}

\vspace*{-0.6cm}
\section{Conclusions \label{sum}}
The morphology and 2D color patterns of high-\zet\ starburst galaxies (and AGN) can be deceptive.
This is because they are ObsF projections of a \emph{spatially inhomogeneous} optical-UV SED that, depending on \zet\ and the
filters a galaxy is observed with, can markedly differ from its true (rest-frame) characteristics.
The root of the problem, dubbed \cmod\ (chromatic surface brightness modulation; \tref{P23}), lies in the differential
dimming of UV-faint (old, passively evolving or dusty) regions in combination with the brightening of UV-emitting (young SF) regions.
Additional elements shaping the spatially and time evolving 2D SED of a starburst galaxy, and thus contributing to \cmod, are
a) strong nebular (line and continuum) emission
and b) its spatial decoupling from ionizing YSCs as a manifestation of SF feedback and the reprocessing of locally leaking \lyc\ radiation
into \ne\ hundreds of pc away from the locus where it is generated, and c) frequently differing extinction patterns in the stellar and nebular component.

Using the nearby BCG \object{Haro 11} as an example, we show that \cmod\ drastically affects the morphology and color patterns of
higher-\zet\ starburst galaxies observed with the JWST and Euclid. Color maps of such systems, if taken at face value (i.e., uncorrected for \cmod\ effects) unavoidably lead to a broad range of erroneous conclusions about the nature and evolutionary status of such systems.
A physical characterization of higher-\zet\ starburst galaxies from color maps requires spatially resolved \kc-corrections. Since these in turn require a reasonable guess on the rest-frame 2D SED from photometric data they pose a non-trivial task, especially when the redshift of a galaxy is unknown.

The goal of this study is to invite the community to a joint exploration of techniques for a better understanding and the rectification of \cmod\ effects. This is a key prerequisite for fully unleashing the potential of JWST and Euclid for elucidating the starburst phenomenon and its
rôle in the cosmic scenery.

\vspace*{-0.2cm}
\begin{acknowledgements}
We are thankful to I. Breda, A. Adamo, J. Afonso, A. Bik, J. Brinchmann, M. Heyes, A. Paulino-Afonso, E. Rivera-Thorsen and M. Sirressi for stimulating discussions and valuable comments.
PP gratefully acknowledges support by the Wenner-Gren Foundation and the hospitality of the Astronomy Department at Stockholm University, as well as Funda\c{c}\~{a}o para a Ci\^{e}ncia e a Tecnologia (FCT) grants UID/FIS/04434/2019, UIDB/04434/2020, UIDP/04434/2020 and Principal Investigator contract CIAAUP-092023-CTTI. G\"O acknowledges support from the Swedish Research Council (VR) and the Swedish National Space Administration (SNSA).
\end{acknowledgements}

\vspace*{-0.8cm}

\pagebreak
\begin{appendix}
\section{Supplementary notes \label{app}}
\subsection{Spectral fitting\label{app:SFH}}
Supplementing the discussion in Sect.~\ref{IFS} we show the results from fitting the spectrum of region A--C with the population spectral synthesis code \starlight\ \citep{Cid05} after bidimensional subtraction of nebular continuum emission from the IFS data cube. The spectral modeling was carried out with the same library of simple stellar population (SSP) spectra as the one used in \tref{P23}, comprising 236 templates from \cite{BC03} for a Chabrier initial mass function that cover an age between 1 Myr and 13.5 Gyr and a metallicity between \zsun/50 and 2.5\,\zsun.
Spectral fits were computed in two setups, the first assuming extinction after \citet{Cardelli89} and the second the \citet{Calzetti00} attenuation law (CCM and CAL, respectively), finding a good mutual agreement.

\begin{center}
\begin{figure}
\begin{picture}(200,224)
\put(20,150){\includegraphics[width=6.1cm]{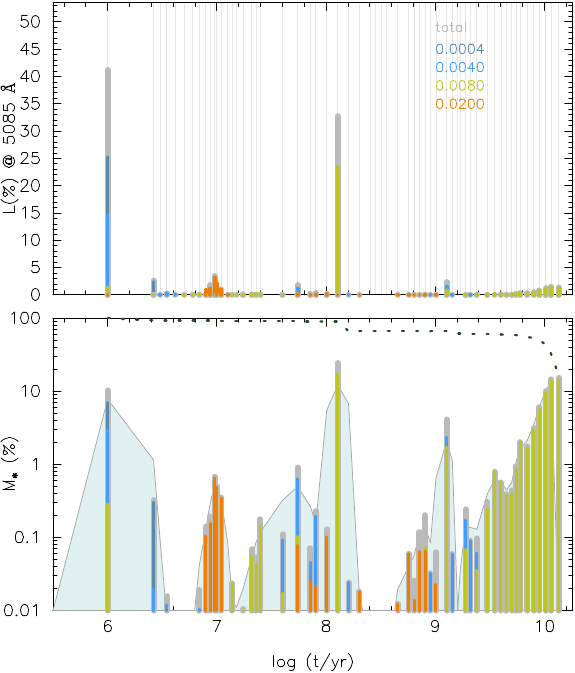}}
\put(20,75){\includegraphics[width=6.1cm]{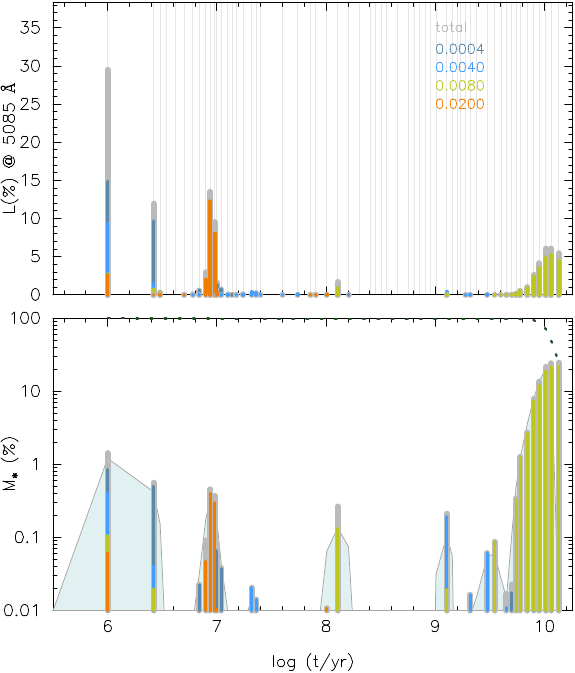}}
\put(20,0){\includegraphics[width=6.1cm]{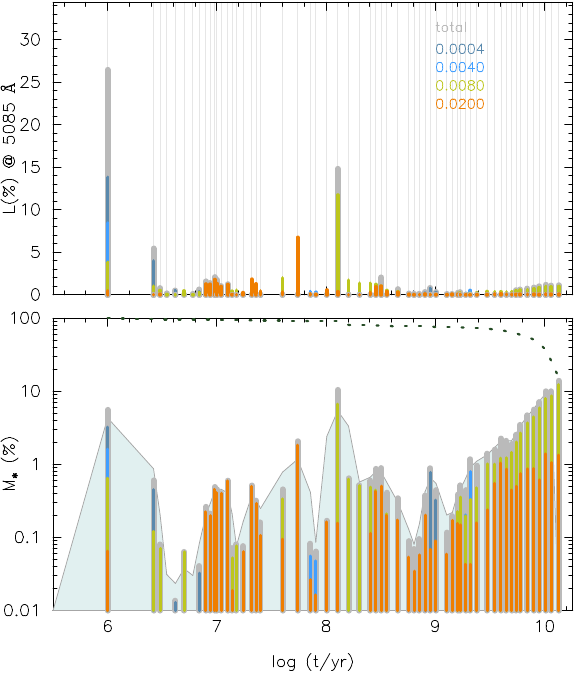}}
\PutLabel{28}{217.4}{\nlx \textcolor{black}{region A}}
\PutLabel{28}{142}{\nlx \textcolor{black}{region B}}
\PutLabel{28}{67.5}{\nlx \textcolor{black}{region C}}
\end{picture}
\caption{Best-fitting population vectors for regions A--C obtained with \starlight. The luminosity and mass contribution (in \%) of SSPs of different age and metallicity (cf. color coding on the upper-right) is shown in the upper and lower panels, respectively.
Gray vertical lines mark the age of the library SSPs and the dotted curve shows the cumulative growth of the stellar mass \mstar.}
\label{app:SFHs}
\end{figure}
\end{center}

As apparent from Fig.~\ref{app:SFHs}, star formation in region A has significantly increased since $\sim$100 Myr whereas that in region C has kept a relatively constant level over the past $\sim$1 Gyr. It can be seen that the stellar mass in all three SF regions is dominated by the underlying old ($>$3 Gyr) stellar host. A detailed surface photometry study of the latter by \cite{BO02} permitted determination of its average color to $B$-$V$=0.87 mag and $B$-$R$=1.24 mag, establishing that starburst activity in \object{Haro 11} takes place in the central part of an evolved stellar component.

\subsection{Simulations of \cmod\ effects \label{app:neb}}
As described in \tref{P23}, the panchromatic UV-through-IR rest-frame SED $F_0$ at each spaxel (or segment of binned spaxels) was computed
from the best-fitting population vector obtained with \starlight, and after addition of nebular continuum and line emission.
The resulting composite SED was in turn simulated in the redshift range $0\leq z \leq 5.4$ as $F(\lambda,z)=F_0(\lambda/(1+z))/(1+z)$ and convolved with filter transmission curves to compute ObsF magnitudes in the filters $UBVRIJHK$ (Vega system) and JWST F070W, F090W, F115W, F150W, F200W, F277W, F444W, F152M, F360M, F410M, as well as Euclid VIS, Y, J, H (AB system). These values, referred to in \tref{P23} as \emph{reduced} surface brightness $\mu\arcmin$, take only bandpass shift and wavelength stretching into account. Cosmological surface brightness dimming
was neglected since it equally impacts all galaxy structural components (SF regions and the old underlying stellar host), it is thus
unimportant for variation across \zet\ of the morphology and the color maps, which is the focus of this study.

\begin{center}
\begin{figure}
\begin{picture}(86,214)
\put(5,194){\includegraphics[width=8cm]{fig/zcol03_F090W_F115W.pdf}}
\put(5,175){\includegraphics[width=8cm]{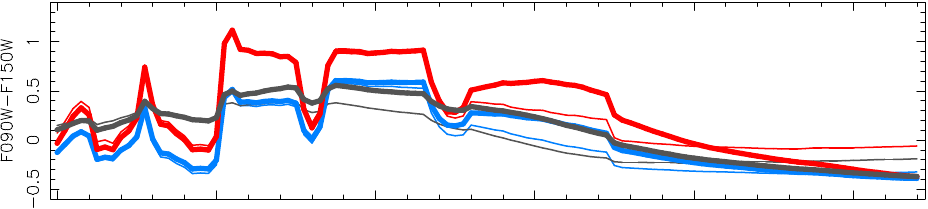}}
\put(5,156){\includegraphics[width=8cm]{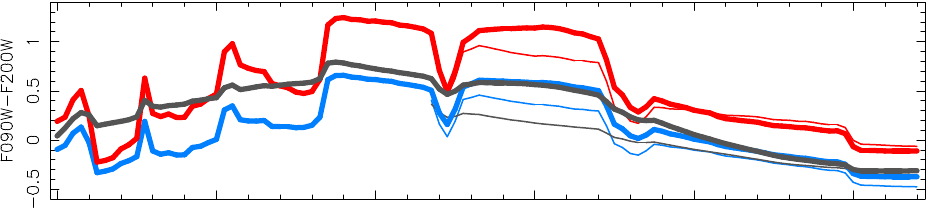}}
\put(5,137){\includegraphics[width=8cm]{fig/zcol04_F115W_F150W.pdf}}
\put(5,118){\includegraphics[width=8cm]{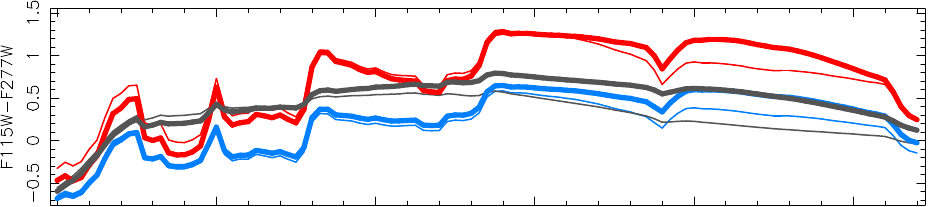}}
\put(5,99){\includegraphics[width=8cm]{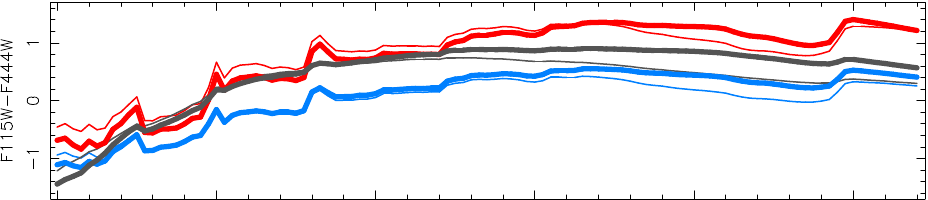}}
\put(5,80){\includegraphics[width=8cm]{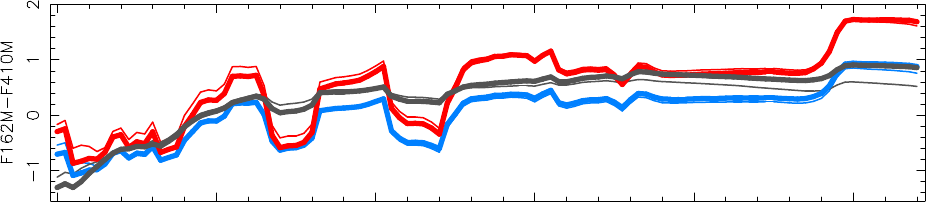}}
\put(5,61){\includegraphics[width=8cm]{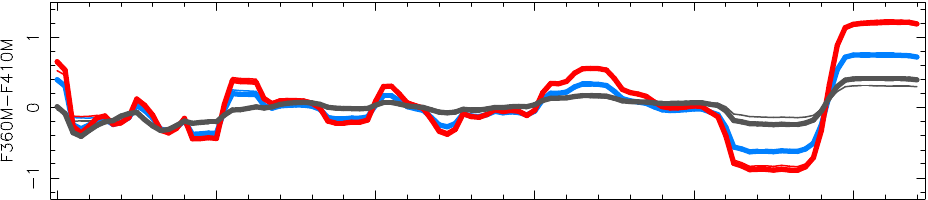}}
\put(5,42){\includegraphics[width=8cm]{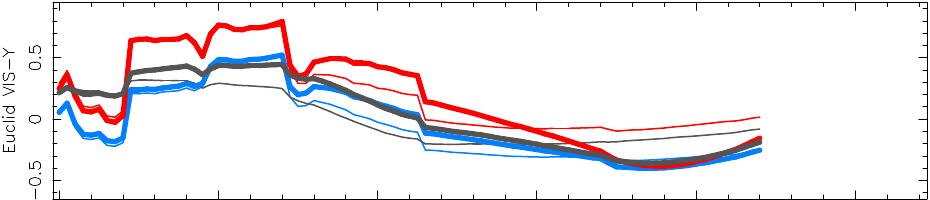}}
\put(5,22){\includegraphics[width=8cm]{fig/zcol11_EuclidY_H.pdf}}
\put(5,0){\includegraphics[width=8cm]{fig/zcol12_EuclidJ_H.pdf}}
\end{picture}
\caption{ObsF color of regions A-C vs. redshift. Solid and dashed curves refer, respectively, to the \citet[][CCM]{Cardelli89} extinction law and the \citet[][CAL]{Calzetti00} attenuation law.
\label{app:col_vs_z}}
\end{figure}
\end{center}

Nebular emission lines outside the observed spectral range were computed based on theoretical flux ratios relative to H$\beta$, following prescriptions encoded in the evolutionary synthesis model \pegase~2 \citep{FRV97}. The extinction applied to the nebular (line and continuum) emission in this spectral interval was based on the observed H$\alpha$/H$\beta$ ratio.
Even though line ratios in high-\zet\ protogalaxies assembling their first stellar population out of metal-free gas may appreciably differ from typical values in the local universe, inter alia, because of the contribution of shocks \citep{Brinchmann23} and the high ionization parameter $U$ expected for these sources, we consider that Fig.~\ref{col_vs_z} offers a reasonable first-order approximation to the ObsF color for a high-\zet\ analog of \object{Haro 11}. At the same time, it is important ro remark that our simulations are not evolutionary consistent, that is, do not take into account the youthening of stellar populations (specifically, of the underlying host) with increasing \zet\ (see \tref{P23} for a discussion of this subject and further simulations referring to evolutionary consistent models for parametric star formation histories) and do not incorporate any prescription for the chemical evolution of stars and gas. This would require several uncertain assumptions on a number of crucial and closely interlinked elements of galaxy evolution, such as infall of metal-poor gas from the cosmic web and ejection of metal-enriched gas generated in starburst episodes, dispersal and mixing of heavy elements within a galaxy, the metallicity-dependent cooling efficiency of gas and the timescales for the reincorporation of heavy elements into subsequent stellar generations, and a possible dependence of the IMF on age, gas-phase metallicity and SF stochasticity.
Clearly, a probably density-bound EELG with an [O{\sc iii}]$_{5007}$/[O{\sc ii}]$_{3727,29}$ (O$_{32}$) ratio as high as $\sim$53 and an EW(H$\beta$)$\sim$577 \AA\ like \object{J2229+2725} \citep{Izotov21-J2229+2725} will follow a different trajectory in Figs.~\ref{col_vs_z} and \ref{app:col_vs_z} than a low-$U$ metal-rich spiral galaxy, or a centrally dust-obscurred ultra-luminous IR galaxy (ULRIG) like \object{Arp 220}.
It also should be kept in mind that nebular emission in \object{Haro 11} is excited by relatively metal-rich stellar populations that certainly are not representative of low-mass protogalaxies (our spectral fits yield a light-weighted stellar metallicity of $\sim$0.3, 0.4 and 0.5 \zsun, for A, B and C, respectively) and that the SED of the three SF knots that was used as input for the simulations in Fig.~\ref{col_vs_z} has not been decontaminated from the low yet non-negligible light contribution from stars composing the underlying old stellar host.
\begin{center}
\begin{figure}
\begin{picture}(86,77)
\put(2,0){\includegraphics[width=8.7cm, angle=0]{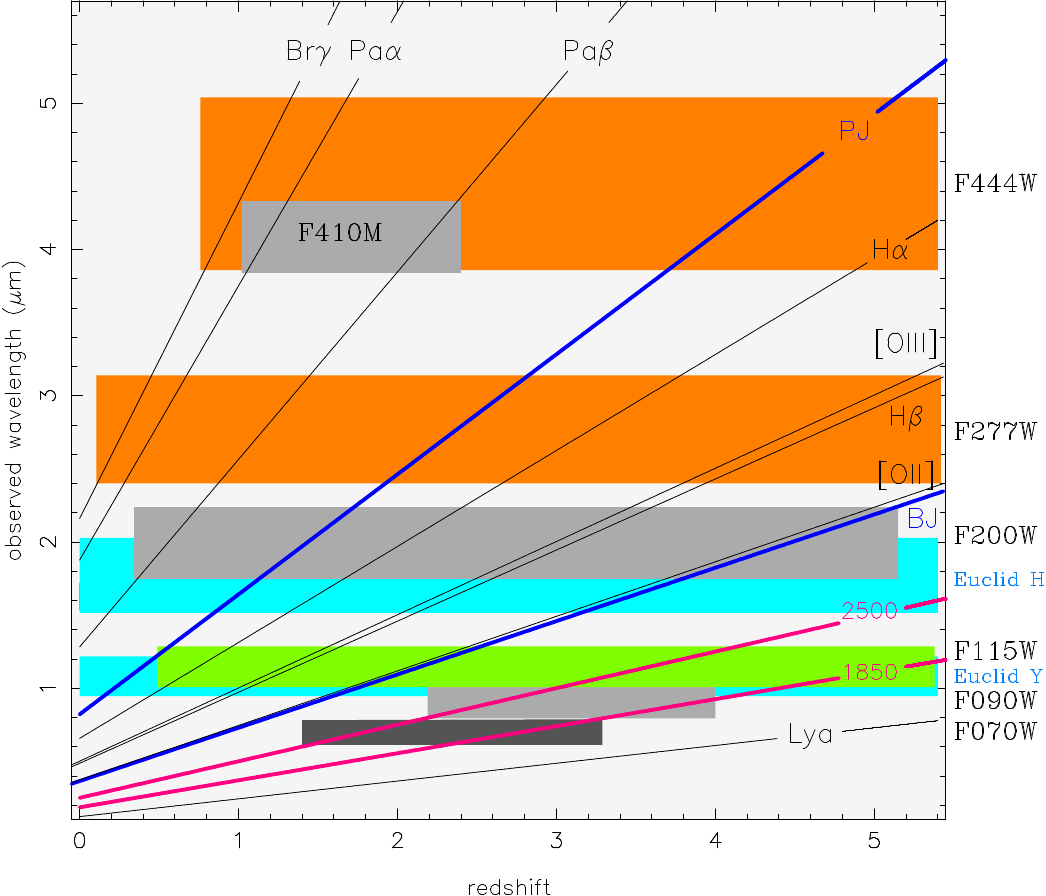}}
\end{picture}
\caption{ObsF wavelength of some emission lines as a function of redshift. Additionally, blue lines show the ObsF wavelength of the Balmer and Paschen jump (BJ and PJ, respectively) and red lines bracket the restframe wavelength range between 1850 \AA\ and 2500 \AA\ around the 2175 \AA\ absorption bump. At intermediate redshifts (1.4$\la z \la$4.2) the latter mostly impacts
photometry in the filters F070W, F090W, F115W and Euclid Y, whereas filters with a longer $\lambda_0$ (e.g., Euclid~H, F200W and F277W) are immune to it. At the same time, photometry in these filters can be affected by the BJ and optical emission lines such as [O{\sc ii}] and \ha+[N{\sc ii}].}
\label{zetfil}
\end{figure}
\end{center}
\begin{center}
\begin{figure*}
\begin{picture}(200,200)
\put(20,164){\includegraphics[width=7cm]{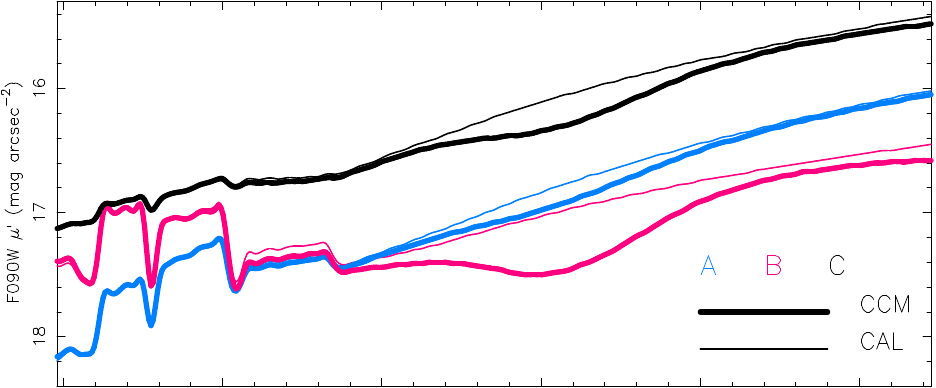}}
\put(20,132){\includegraphics[width=7cm]{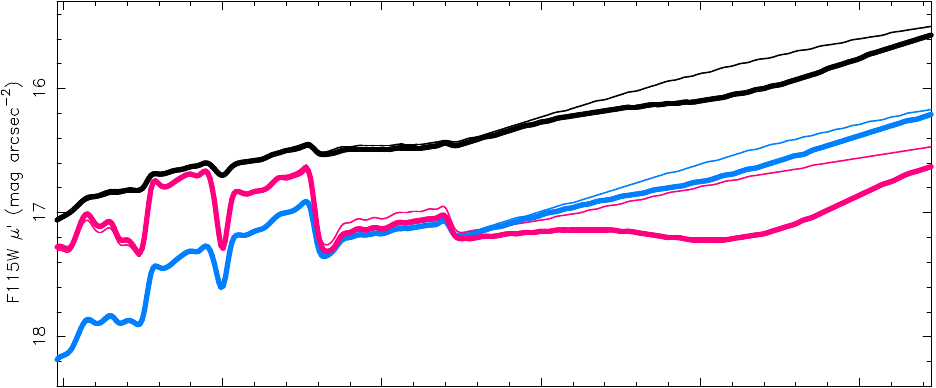}}
\put(20,100){\includegraphics[width=7cm]{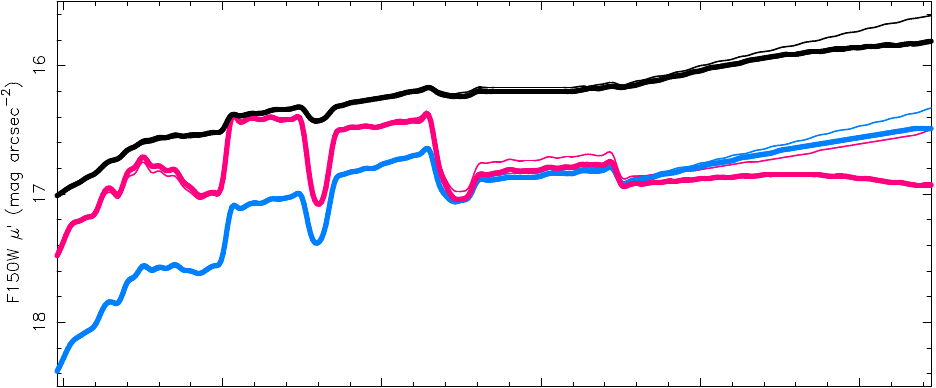}}
\put(20,68){\includegraphics[width=7cm]{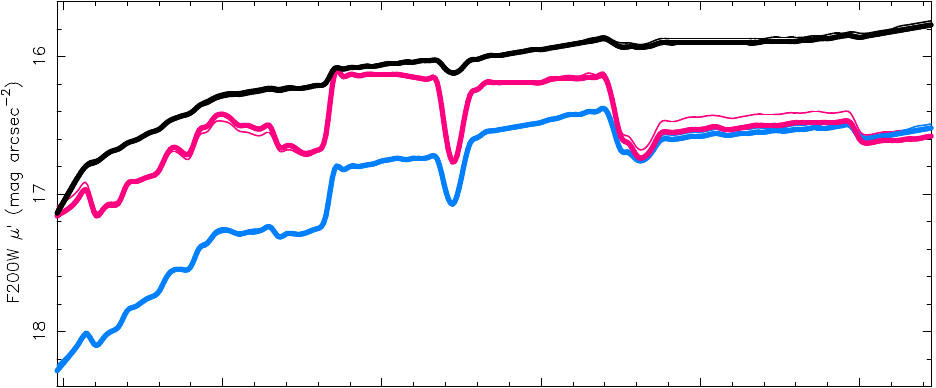}}
\put(20,35){\includegraphics[width=7cm]{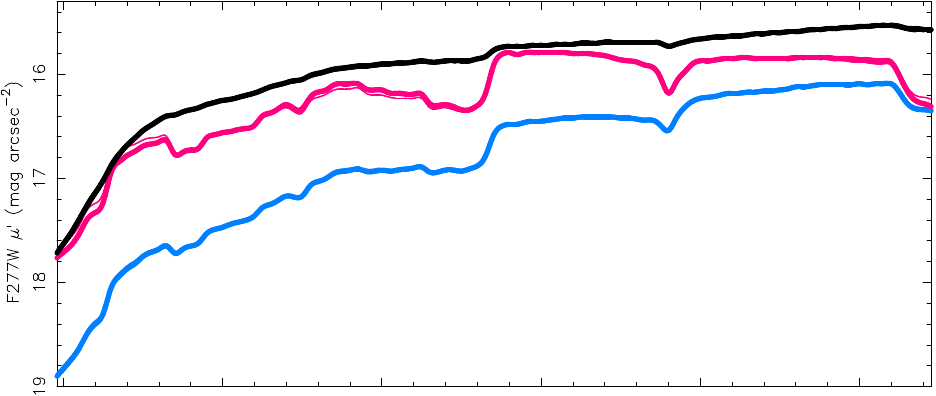}}
\put(20,0){\includegraphics[width=7cm]{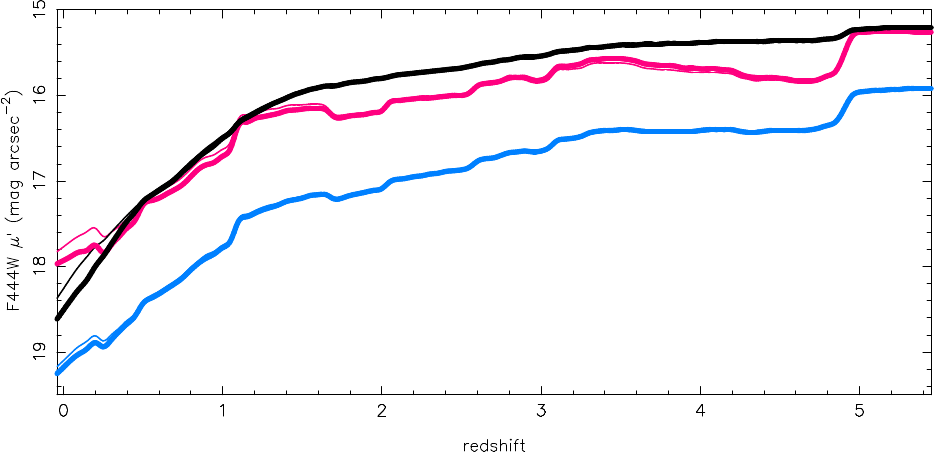}}
\put(100,164){\includegraphics[width=7cm]{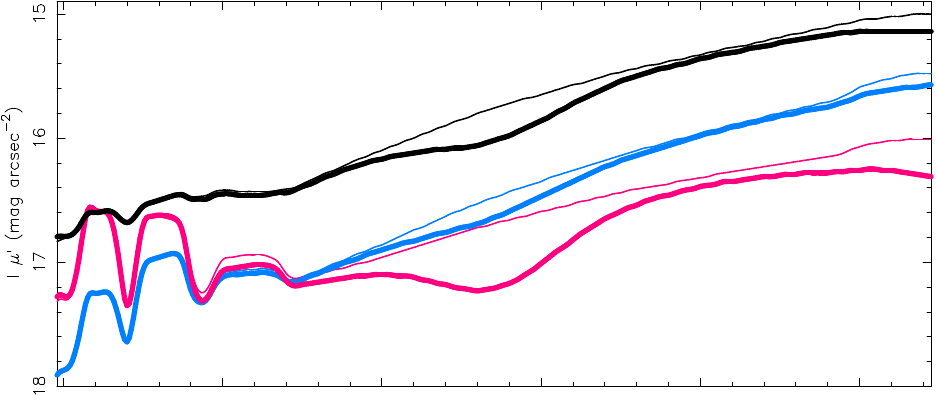}}
\put(100,132){\includegraphics[width=7cm]{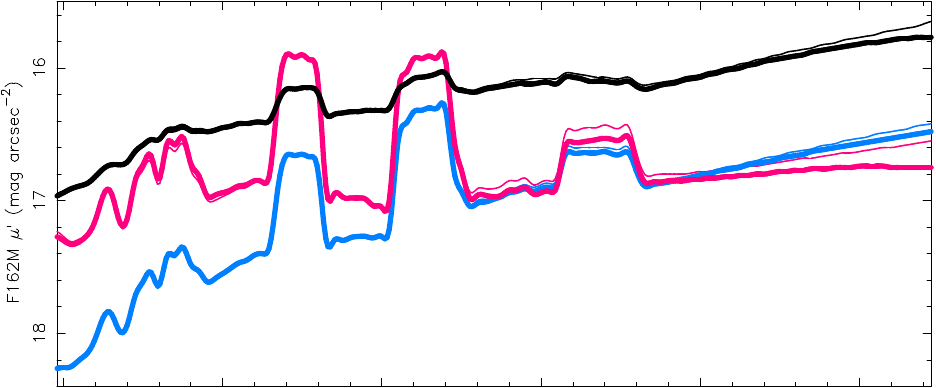}}
\put(100,100){\includegraphics[width=7cm]{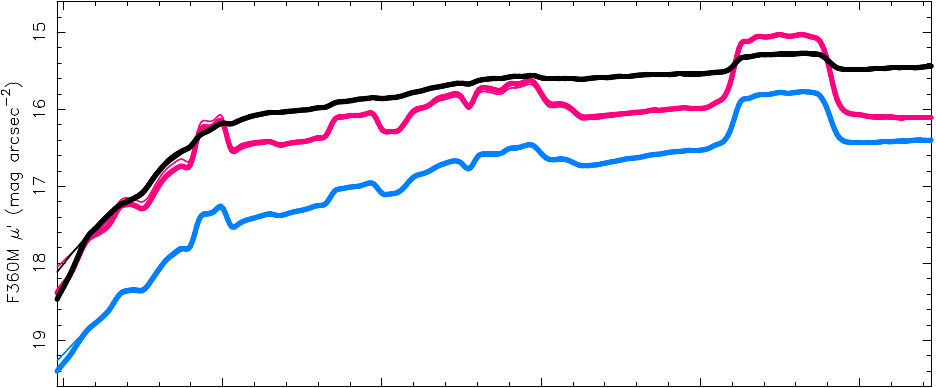}}
\put(100,68){\includegraphics[width=7cm]{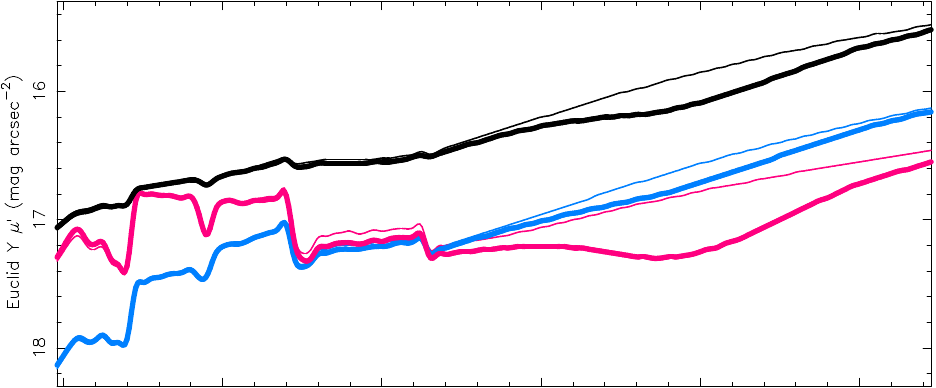}}
\put(100,35){\includegraphics[width=7cm]{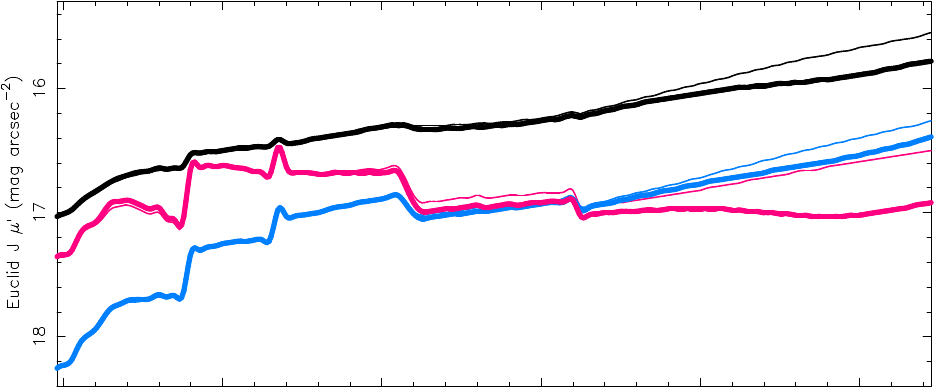}}
\put(100,0){\includegraphics[width=7cm]{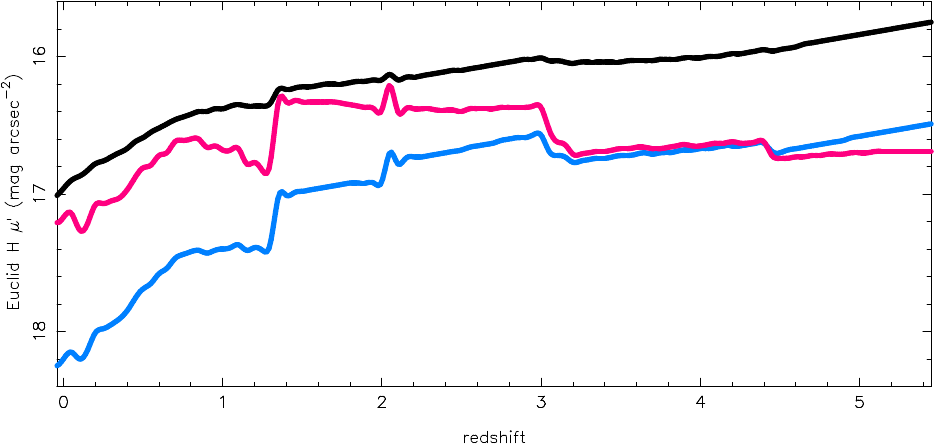}}
\end{picture}
\caption{Variation of the \emph{reduced} surface brightness $\mu\arcmin$ (\sbb) of regions A--C in \object{Haro 11}
as a function redshift for various JWST filters (F090W, F115W, F150W, F200W, F277W, F444W, F162M, F360M; AB system),
the Johnson-Cousins $I$ filter (Vega system), and the Euclid Y, J and H filters (AB system).
Simulations assuming CCM and CAL are shown with, respectively, thick and thin curves.
\label{app:mmag}}
\end{figure*}
\end{center}
From Figs.~\ref{col_vs_z} and \ref{app:col_vs_z} it also can be seen that even a small variation in redshift can lead to a change of the color by $>$0.5 dex, in some cases with a reversion from negative to positive values, depending on the color index considered. This is primarily the result of strong emission lines with a rest-frame EW$\sim 10^3$ \AA\ (close to the effective width of broadband filters) and other nebular features (e.g., the Balmer and Paschen jump at 3646 \AA\ and 8207 \AA, respectively) moving in and out of filter transmission curves.
The fact these have a relatively sharp boundary for most filters onboard JWST and all filters onboard Euclid  (contrary to, e.g., Johnson/Bessel UBVRI filters whose transmission varies relatively smoothly across $\lambda$) amplifies these strong local color discontinuities.
Thus, substantial color differences between EELGs with a nearly identical rest-frame SED yet a slightly different redshift are to be expected from JWST and Euclid photometry.

The origin of the color discontinuities can better be understood with the help of Fig.~\ref{zetfil} (and its comparison with Figs.~\ref{dmag}, \ref{app:mmag} and \ref{app:col_vs_z}) that depicts redshift intervals where strong nebular features, and additionally the broad $\lambda$ interval of enhanced intrinsic extinction around the 2175 \AA\ bump (between $\sim$1850 \AA\ and $\sim$2500 \AA) fall within different filters.
For example, the sudden reddening by 1 mag of region B at $1.25\la z \la 1.5$ in the Euclid Y-H color is because the strong \ha+[N{\sc ii}] lines move into the H band, whereas the rest-frame interval of $\sim$2100--3600 \AA\ covered by the Y filter does not contain strong emission lines.
Likewise, the brightening of $\mu\arcmin$(F444W) at \zet$\sim$4.7 (Fig.~\ref{app:mmag}) with the simultaneous reddening of the F115W-F444W color
(Fig.~\ref{app:col_vs_z}) echoes the shift of \ha+[N{\sc ii}] into the F444W filter. From our foregoing remarks it follows that a combined analysis of spatially resolved colors in JWST or Euclid filters, in particular, the search for seemingly discordant combinations of color indices offers an efficient means for the identification and redshift redermination of distant EELGs, and eventually allows better isolating Ly$\alpha$ emitters from non-Ly$\alpha$ emitting galaxies with a similar rest-frame optical SED. In this context, the core-to-envelope color contrast is a far more sensitive diagnostic for EELGs with spatially extended nebular emission than the integral (luminosity-weighted) colors of such systems (\tref{PO12}).

Finally, a crucial yet poorly constrained factor is the 2D intrinsic extinction (more specifically, attenuation) that the rest-frame UV SED of a  high-\zet\ starburst galaxy suffers along different sightlines and in its stellar and gaseous constituent. The 2175 \AA\ bump, prominent in the case of the CCM extinction law whereas absent in the case of the CAL attenuation law, is an important unknown in this regard.
This is apparent from Fig.~\ref{col_vs_z} (and the extended set of colors in Fig.~\ref{app:col_vs_z}) showing that the choice between CCM or CAL
translates into differences of up to $\sim$0.3 mag in the ObsF colors for certain combinations of filters when the intrinsic visual extinction is significant (0.5-1 $V$ mag), as in region B: whereas CCM yields at \zet=3 for this region by $\sim$0.3 mag redder F090-F115W and F090W-F200W colors than CAL (because the 2175 \AA\ bump affects the ObsF SED at $\sim$0.87 $\mu$m, i.e., within the F090W filter and outside the F115W and F200W filters),it implies at \zet=4.4 a by $\sim$0.3 mag bluer F090-F115W color than CAL, as it only impacts the F115W filter.
For similar reasons, CCM implies for region B a by 0.3 mag redder Euclid Y-H (J-H) color at \zet=4 (5) than CAL. As previously pointed out (Fig.~\ref{dmag}), intrinsic UV extinction is also an important determinant of the ObsF morphology of high-\zet\ starburst galaxies.
For example, the F115W C/B ratio for \object{Haro 11} at \zet$\sim$4.4 will increase from $\sim$2.4 to $\sim$3 if CCM is assumed instead of CAL.

\subsection{Stellar vs. nebular extinction and \cmod\ effects: further illustrative examples \label{app:synopsis}}
This section provides a visual impression of the variation of the morphology and color patterns of star-forming galaxies across redshift based on simulations of \cmod\ effects following \citet{P23}. Additionally, it is meant to illustrate the presence of extended nebular emission in many star-forming galaxies, as well as differences and similarities in their stellar and nebular extinction. With the exception of the BCGs \object{Haro 11} and \object{ESO 338-IG04} \citep{BO02}, for which own data with the IFS unit MUSE are available (PI: \"Ostlin), the MUSE IFS data used for spectral modeling with \starlight\ and subsequently for the simulation of \cmod\ effects were retrieved in reduced form from the ESO Science Archive ({\tt archive.eso.org}).
The following synopsis includes, besides \object{Haro 11} and \object{ESO 338-IG04}, the BCG
\object{He 2-10} for which MUSE data were acquired in the ESO observig program (OP) ESO 095.B-0321A (PI: Vanzi).
Additionally, examples of more massive star-forming galaxies with different morphologies and a higher intrinsic dust obscuration, some
of those undergoing interaction-induced starburst activity, include \object{IC 1623} (\object{VV 114}; OP: 0100.B-0116A, PI: Carollo),
\object{NGC 7252} (OP: 099.B-0281A; PI: Privon), \object{II Zw 96} (OP: 097.B-0427A; PI: Privon), \object{Cartwheel galaxy} (OP: 60.A-9333A; science verification) and the ULIRG \object{Arp 220} (OP: 0103.B-0391A; PI: Arribas).
Examples of spiral galaxies, with main focus on the nuclear region of these systems, are given through
\object{NGC 1097} (OP: 097.B-0640A; PI: Gadotti),
\object{NGC 1300} (OP: 097.B-0640A; PI: Gadotti),
\object{NGC 1365} (OP: 094.B-0321A; PI: Marconi),
\object{NGC 2775} (OP: 0104.B-0404A; PI: Erwin),
\object{NGC 3351} (\object{M95}; OP: 097.B-0640A; PI: Gadotti),
\object{NGC 4045} (OP: 0101.A-0282A; PI: Contini),
\object{IC 2051} (OP:0104.B-0404A; PI: Erwin), and
\object{ESO 498-G05} (OP: 096.B-0309A; PI: Carollo).
Finally, the case of extended AGN-driven nebular emission is exemplified through
\object{NGC 5972} (OP: 0102.B-0107A; PI: Sartori) and \object{Teacup} (OP: 0102.B-0107A; PI: Sartori).

Animations of \cmod\ effects out to \zet=5.4 in steps of 0.05 can be downloaded in m4a format from {\color{red}\tt http://hyperlink:tbd},
and data cubes in FITS format can be provided upon request.

\imCMOD{haro11SUnoNeb}{Haro 11 (D=82 Mpc). Supplementing the material in Sect.~\ref{IFS}, the top-left panel shows the \ha+[N{\sc ii}]
map of the BCG, with the \ewha\ and gas velocity map within the depicted central area of the galaxy shown, respectively, in the top-middle and top-right panel. The reader is referred to \tref{Menacho et al. 2019 (MNRAS, 487, 3183)} for a detailed kinematical study of the nebular component of this BCG.
The lower panels (from top to bottom) show the \emph{reduced} surface brightness $\mu\arcmin$ (\sbb) of the galaxy in the JWST F090W filter, and simulated JWST F090W-F150W and Euclid Y-H color maps (AB system) for \zet=0, 0.15, 0.35, 0.7, 0.9, 1.3, 1.6, 2, 2.5, 3, 4 and 5, and in $V$-$I$ (Vega system) for \zet=0, 0.05, 0.3, 0.4, 0.7, 0.9, 1, 1.3, 1.6, 2, 2.3 and 2.7.}{mycyan}{HaF115W.png}{EWHa.png}{vHa.png}{app:Haro11}{F090W}
\imCMOD{eso338SUnoNeb}{ESO338-IG04 (D=37.5 Mpc; see \tref{Bik et al. 2018, A\&A, 619, A131} for a detailed analysis).
The larger upper-row panels show (from left to right) the \ha+[N{\sc ii}] map, and the stellar and nebular extinction map ($V$ mag) within the
depicted central square area. The meaning of the smaller figures in the four lower panels is same as in Fig.~\ref{app:Haro11}.}{mycyan}{HaF090W.png}{AVstellar.png}{AVneb.png}{app:ESO338}{F090W}
\imCMOD{he210}{He 2-10 ($D$= 8.23 Mpc). The layout is same as in Fig.~\ref{app:ESO338}.}{mycyan}{HaF090W.png}{AVstellar.png}{AVneb.png}{app:He210}{F090W}
\imCMOD{ic1623}{IC 1623 (\object{VV 114}; $D$= 84.4 Mpc). The layout is same as in Fig.~\ref{app:ESO338}.}{mycyan}{HaF090W.png}{AVstellar.png}{AVneb.png}{app:IC1623}{F090W}
\imCMOD{n7252deep}{NGC 7252 ($D$= 66.1 Mpc). The layout is same as in Fig.~\ref{app:ESO338}.}{mycyan}{HaF090W.png}{AVstellar.png}{AVneb.png}{app:NGC7252}{F090W}
\imCMOD{iizw96}{II Zw 96 ($D$= 155.1 Mpc). The layout is same as in Fig.~\ref{app:ESO338}.}{mycyan}{HaF090W.png}{AVstellar.png}{AVneb.png}{app:IIZw96}{F090W}
\imCMOD{cartwheel}{Cartwheel galaxy (D=129.6 Mpc). The layout is same as in Fig.~\ref{app:ESO338}.}{mycyan}{HaF090W.png}{AVstellar.png}{AVneb.png}{app:Cartwheel}{F090W}
\imCMOD{arp220}{Arp 220 (D=83.1 Mpc). The layout is same as in Fig.~\ref{app:ESO338} with the only difference being that the $\mu\arcmin$ map (second row) refers to the JWST F150W filter.}{mycyan}{HaF090W.png}{AVstellar.png}{AVneb.png}{app:Arp220}{F150W}
\begin{center}
\begin{figure*}
\begin{picture}(200,60)
\put(0,0){\includegraphics[width=8.0cm]{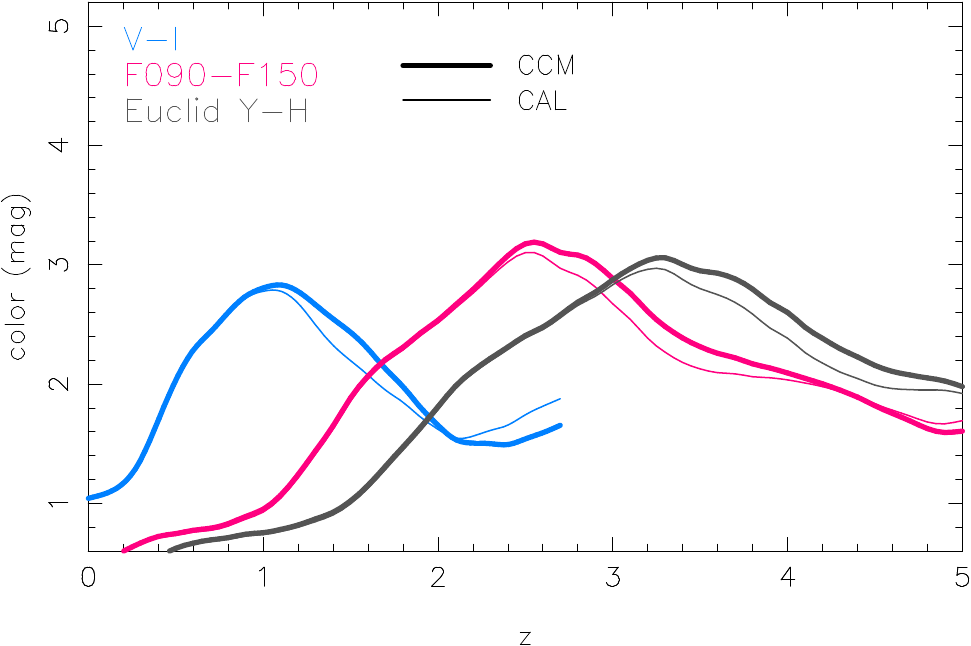}}
\put(100,0){\includegraphics[width=8.0cm]{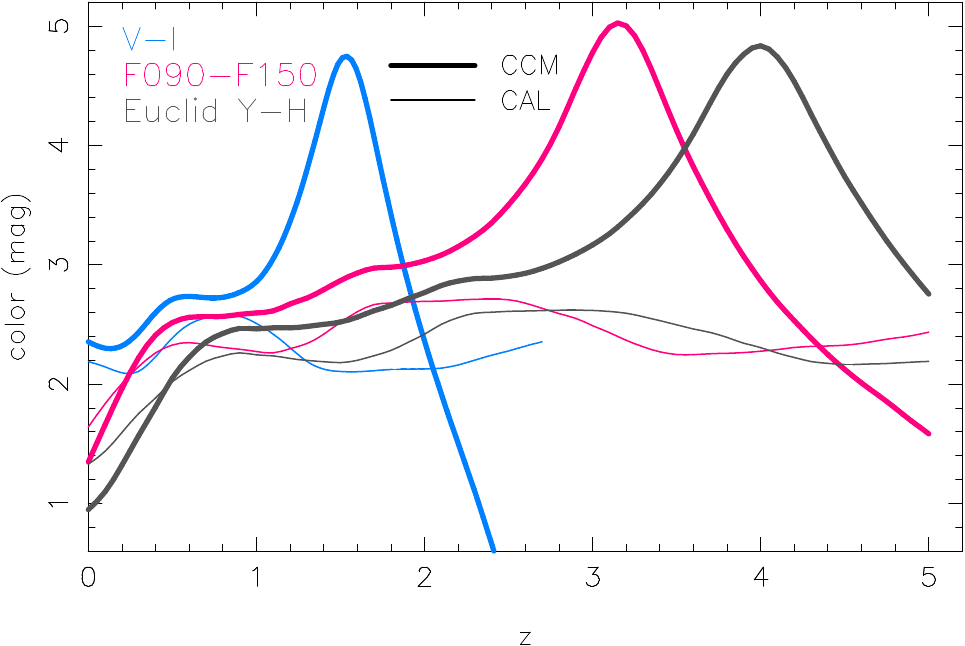}}
\end{picture}
\caption{Illustration of the effect of intrinsic extinction on the predicted $V$-$I$, F090W-F150W and Euclid Y-H color as a function of redshift
in the case of CCM and CAL (thick and thin curves, respectively).
The l.h.s. panel refers to a randomly selected spaxel in the low-obscuration periphery of \object{Arp 220} where nebular emission is very weak (\ewha$\sim$5 \AA), and the r.h.s. panel to the central highly obscurred region (\exstars=4 mag and \exneb=6.2 mag) where nebular emission is modest (\ewha$\sim$43 \AA). In the latter case, intrinsic extinction after CCM leads to a reversion by $\sim$2 mag of $V$--$I$, JWST F090W-F150W and Euclid Y-H colors at, respectively, \zet$\sim$1.5, 3.2 and 4. It can be seen that the $V$-$I$ color at \zet$\sim$2.4 and the F090W-F150W color at \zet$\sim$ become bluer than the value predicted by CAL.
Thus, a seeming paradox in the case of CCM and for certain colors is that highly obscured regions in a distant galaxy appear relatively blue, with the  effect becoming stronger the higher the intrinsic extinction is.
\label{app:spaxels}}
\end{figure*}
\end{center}
\imCMOD{n1097}{NGC 1097 ($D$= 22.7 Mpc; see \tref{Gadotti et al. 2018, MNRAS, 482, 506} for details). The layout is same as in Fig.~\ref{app:ESO338}.
}{mycyan}{HaF090W.png}{AVstellar.png}{AVneb.png}{app:NGC1097}{F090W}
\imCMOD{n1300}{NGC 1300 ($D$= 21.1 Mpc; see \tref{Gadotti et al. 2018} for details). The layout is same as in Fig.~\ref{app:ESO338}.}{mycyan}{HaF090W.png}{AVstellar.png}{AVneb.png}{app:NGC1300}{F090W}
\imCMOD{n1365center}{Central part of NGC1365 ($D$= 22.7 Mpc; see \tref{Gadotti et al. 2018} for details). The layout is same as in Fig.~\ref{app:ESO338}.}{mycyan}{HaF090W.png}{AVstellar.png}{AVneb.png}{app:NGC1365center}{F090W}

\imCMOD{n2775}{NGC 2775 ($D$= 24.5 Mpc). The layout is same as in Fig.~\ref{app:ESO338}.}{mycyan}{HaF090W.png}{AVstellar.png}{AVneb.png}{app:NGC2775}{F090W}

\imCMOD{n3351}{NGC3351 ($D$= 16.6 Mpc). The layout is same as in Fig.~\ref{app:ESO338}.}{mycyan}{HaF090W.png}{AVstellar.png}{AVneb.png}{app:NGC3351}{F090W}
\imCMOD{n4045}{NGC 4045 ($D$= 34.2 Mpc). The layout is same as in Fig.~\ref{app:ESO338}.}{mycyan}{HaF090W.png}{AVstellar.png}{AVneb.png}{app:NGC4045}{F090W}
\imCMOD{ic2051}{IC 2051 ($D$= 26.3 Mpc). The layout is same as in Fig.~\ref{app:ESO338}.}{mycyan}{HaF090W.png}{AVstellar.png}{AVneb.png}{app:IC2051}{F090W}
\imCMOD{eso498g05}{ESO 498-G05 ($D$= 32.8 Mpc). The layout is same as in Fig.~\ref{app:ESO338}.}{mycyan}{HaF090W.png}{AVstellar.png}{AVneb.png}{app:ESO498G05}{F090W}
\imCMOD{n5972}{NGC 5972 ($D$= 132.9 Mpc). The layout is same as in Fig.~\ref{app:ESO338}.}{mycyan}{HaF090W.png}{AVstellar.png}{AVneb.png}{app:NGC5972}{F090W}
\imCMOD{teacup}{Teacup galaxy ($D$= 379 Mpc). The layout is same as in Fig.~\ref{app:ESO338}.}{mycyan}{HaF090W.png}{AVstellar.png}{AVneb.png}{app:Teacup}{F090W}
\end{appendix}

\begin{thebibliography}{}
\baselineskip=0ex
\bibitem[Bacon et al.(2014)]{Bacon14}Bacon,\,R.,\,Vernet,\,J.,\,Borisova,\,E., et al. 2014, The Messenger, 157, 13
\bibitem[Bergvall \& Olofsson(1986)]{BO86}Bergvall, N. \& Olofsson, K. 1986, A\&AS, 64, 469
\bibitem[Bergvall \& \"Ostlin(2002)]{BO02}Bergvall, N. \& \"Ostlin, G., 2002, A\&A, 390, 891
\bibitem[Bergvall et al.(2006)]{Bergvall06}Bergvall, N., et al. 2006, A\&A, 448, 513
\bibitem[Brinchmann(2023)]{Brinchmann23}Brinchmann, J. 2023, MNRAS, 525, 2087
\bibitem[Bruzual \& Charlot(2003)]{BC03} Bruzual, G. \& Charlot, S., 2003, MNRAS, 344, 1000
\bibitem[Calzetti et al.(2000)]{Calzetti00}Calzetti D., Armus L., Bohlin R.C. et al. 2000. ApJ 533, 682
\bibitem[Cardelli et al.(1989)]{Cardelli89}Cardelli J. A., Clayton G. C., Mathis J. S., 1989, ApJ, 345, 245
\bibitem[Cardoso et al.(2016)]{Cardoso16}Cardoso, L.S.M., Gomes, J.M. \& Papaderos 2016, A\&A 594, L2
\bibitem[Cid Fernandes et al.(2005)]{Cid05}Cid\,Fernandes,\,R.,Mateus,\,A.,\,Sodr\'{e},\,L.\,et\,al.,\,2005,\,MNRAS,\,358,\,363
\bibitem[Fioc \& Rocca-Volmerange(1997)]{FRV97}Fioc, M., \& Rocca-Volmerange, B. 1997, A\&A, 326, 950
\bibitem[Gomes \& Papaderos(2017)]{GP17}Gomes, J.M. \& Papaderos, P., 2017, A\&A, 602, A63    
\bibitem[Gross et al.(2021)]{Gross21}Gross, A.C., Prestwich, A. \& Kaaret, P. 2021, MNRAS, 505, 610
\bibitem[Hayes et al.(2007)]{Hayes07}Hayes, M., \"Ostlin, G., Atek, H., Kunth, D. 2007, MNRAS, 382, 1465
\bibitem[Izotov et al.(2021)]{Izotov21-J2229+2725}Izotov, Y.I., Thuan, T.X. \& Guseva, N.G. 2021, MNRAS 504, 3996
\bibitem[Leitet et al.(2011)]{Leitet11}Leitet, E., Bergvall, N., Piskunov, N. et al., 2011, A\&A, 532, A107
\bibitem[Le Reste et al.(2023)]{LeReste23}Le Reste, A., Cannon, J.M., Hayes, M.J. et al. 2023, Nature Astronomy, submitted (arXiv:2301.02676)
\bibitem[Menacho et al.(2019)]{Menacho19}Menacho, V., \"Ostlin, G., Bik, A., et al. 2019, MNRAS, 487, 3183
\bibitem[Menacho et al.(2021)]{Menacho21}Menacho, V., \"Ostlin, G., Bik, A., et al. 2021, MNRAS, 506, 1777
\bibitem[Narayanan et al.(2018)]{Narayanan18}Narayanan, D., Conroy, C., Dav\'e, R. et al. 2018, ApJ, 869, 70
\bibitem[\"Ostlin et al.(2003)]{Ostlin03}\"Ostlin, G., Zackrisson, E., Bergvall, N. et al. 2003, A\&A, 408, 887
\bibitem[\"Ostlin et al.(2009)]{O09}\"Ostlin, G., Hayes, M., Kunth, D., et al., 2009, AJ, 138, 923
\bibitem[\"Ostlin et al.(2021)]{O21}\"Ostlin,\,G.,\,Rivera-Thorsen,\,T.E.,\,Menacho,\,V.\,et\,al.\,2021,\,ApJ,\,912,\,155
\bibitem[\"Ostlin et al.(2015)]{O15}\"Ostlin, G., Marquart, T., Cumming, R., Fathi, K., Bergvall, N., Adamo, A., Amram, P., Hayes, M., 2015, A\&A , 583, 55
\bibitem[Papaderos et al.(2002)]{P02} Papaderos,\,P.,\,Izotov,\,Y.I.,\,Thuan,\,T.X.\,et\,al.\,2002,\,A\&A,\,393,\,461\,\tref{(P02)}
\bibitem[Papaderos \& \"{O}stlin(2012)]{PO12} Papaderos, P. \& \"{O}stlin, G., 2012, A\&A, 537, A126 \tref{(PO12)}
\bibitem[Papaderos et al.(2013)]{P13} Papaderos, P., Gomes, J. \& V\'{i}lchez, J. et al. 2013, A\&A, 555, L1
\bibitem[Papaderos, \"Ostlin \& Breda(2023)]{P23}Papaderos, P., \"Ostlin, G. \& Breda, I. 2023, A\&A 658, A74 \tref{(P23)}
\bibitem[Prestwich et al.(2015)]{Prestwich15}Prestwich, A.H., Jackson, F., Kaaret, P. et al. 2015, ApJ, 812, 166
\bibitem[Reddy et al.(2018)]{Reddy18} Reddy, N.A., Oesch, P.A., Bouwens, R.J. et al. 2018, ApJ, 853, 56
\bibitem[Sirressi et al.(2022)]{Sirressi22}Sirressi, M., Adamo, A., Hayes, M. et al. 2022, MNRAS, 510, 4819
\end{thebibliography}
\end{document}